\documentstyle[preprint,aps]{revtex}
\preprint{Ann.\ Phys.\ (N.Y.), in press (1999)}
\tighten
\draft
\title{The saddle-point method for condensed Bose gases}
\author{Martin Holthaus\footnote{Address after April~1, 1999:
    Ludwig-Maximilians-Universit\"at M\"unchen, Sektion Physik,
    Theresien\-stra{\ss}e~37, D-80333 M\"unchen, Germany}
    and Eva Kalinowski}
\address{Fachbereich Physik der Philipps-Universit\"at,
    Renthof 6, D-35032 Marburg, Germany}
\date{February 25, 1999}
\narrowtext
\newcommand{\oF}{\overline{F}}
\newcommand{\oG}{\overline{G}}
\newcommand{\oH}{\overline{H}}
\newcommand{\of}{\overline{f}}
\newcommand{\eb}{\overline{\eta}}
\newcommand{\Zsp}{Z_N^{(\rm{s.p.})}}
\newcommand{\tZsp}{\widetilde{Z}_N^{(\rm{s.p.})}}
\newcommand{\Isp}{I_{\sigma}^{(\rm{s.p.})}}
\newcommand{\nzcn}{\langle n_{0} \rangle_{\rm{cn}}}
\newcommand{\nocn}{\langle n_{1} \rangle_{\rm{cn}}}
\newcommand{\nacn}{\langle n_{\alpha} \rangle_{\rm{cn}}}
\newcommand{\nncn}{\langle n_{\nu} \rangle_{\rm{cn}}}
\newcommand{\Nex}{\langle N_{\rm ex} \rangle_{\rm cn}}
\newcommand{\fzcn}{\langle \delta^2 n_{0} \rangle_{\rm{cn}}}
\newcommand{\focn}{\langle \delta^2 n_{1} \rangle_{\rm{cn}}}
\newcommand{\facn}{\langle \delta^2 n_{\alpha} \rangle_{\rm{cn}}}
\newcommand{\fncn}{\langle \delta^2 n_{\nu} \rangle_{\rm{cn}}}
\newcommand{\mzcn}{\langle \delta n_{0} \rangle_{\rm{cn}}}
\newcommand{\mocn}{\langle \delta n_{1} \rangle_{\rm{cn}}}
\newcommand{\nzmc}{\langle n_{0} \rangle_{\rm{mc}}}
\newcommand{\namc}{\langle n_{\alpha} \rangle_{\rm{mc}}}
\newcommand{\fzmc}{\langle \delta^2 n_{0} \rangle_{\rm{mc}}}
\newcommand{\famc}{\langle \delta^2 n_{\alpha} \rangle_{\rm{mc}}}
\newcommand{\mzmc}{\langle \delta n_{0} \rangle_{\rm{mc}}}
\newcommand{\nzgc}{\langle n_{0} \rangle_{\rm{gc}}}

\begin{document}

\maketitle

\begin{abstract}
The application of the conventional saddle-point approximation to condensed
Bose gases is thwarted by the approach of the saddle-point to the ground-state
singularity of the grand canonical partition function. We develop and
test a variant of the saddle-point method which takes proper care of this
complication, and provides accurate, flexible, and computationally efficient
access to both canonical and microcanonical statistics. Remarkably, the
error committed when naively employing the conventional approximation in the
condensate regime turns out to be universal, that is, independent of the
system's single-particle spectrum. The new scheme is able to cover all
temperatures, including the critical temperature interval that marks the onset
of Bose--Einstein condensation, and reveals in analytical detail how this
onset leads to sharp features in gases with a fixed number of particles.
In particular, within the canonical ensemble the crossover from the
high-temperature asymptotics to the condensate regime occurs in an
error-function-like manner; this error function reduces to a step function
when the particle number becomes large. Our saddle-point formulas for
occupation numbers and their fluctuations, verified by numerical calculations,
clearly bring out the special role played by the ground state.
\end{abstract}

\pacs{PACS numbers: 05.30.Ch, 05.30.Jp, 03.75.Fi}

\section{Introduction}
\label{S1}

The saddle-point method is one of the most essential tools in statistical
physics~\cite{DarwinFowler23,Schrodinger46}. When comparing different
statistical ensembles, it is used with overwhelming success
in both fundamental theoretical considerations and practical
calculations~\cite{Huang63,Pathria85}. Yet, the conventional form of this
usually easy-to-handle approximation fails in the case of condensed ideal
Bose gases~\cite{Schubert46,FujiwaraEtAl70,ZiffEtAl77}; for instance,
it does not yield the correct fluctuation of the number of condensate
particles~\cite{GajdaRzazewski97,GH97a}.

To elucidate the reason for this failure we consider an ideal Bose gas with
single-particle energies $\varepsilon_{\nu}$, with $\nu = 0,1,2,\ldots$
labeling the individual energy eigenvalues. Since the grand canonical
partition function $\Xi(\beta,z)$ generates the canonical partition
functions $Z_N(\beta)$ by means of the expansion
\begin{eqnarray}
    \Xi(\beta,z) & = &
    \prod_{\nu=0}^{\infty} \frac{1}{1 - z\exp(-\beta\varepsilon_{\nu})}
\nonumber \\
    & = & \sum_{N=0}^{\infty} z^N Z_N(\beta) \; ,
\label{GPF}
\end{eqnarray}
each $N$-particle partition function $Z_N(\beta)$ can be represented,
according to Cauchy's theorem, by a contour integral in the complex $z$-plane,
\begin{equation}
    Z_N(\beta) = \frac{1}{2\pi i} \oint \! {\rm d}z \,
    \frac{\Xi(\beta,z)}{z^{N+1}} \; ,
\label{ZIN}
\end{equation}   
where the path of integration encircles the origin counter-clockwise. As
usual, $\beta = 1/(k_BT)$ is the inverse temperature. Denoting the negative
logarithm of the integrand as $\oF(z)$, i.e., writing 
\begin{equation}
    \frac{1}{z^{N+1}}
    \prod_{\nu=0}^{\infty} \frac{1}{1 - z\exp(-\beta\varepsilon_{\nu})}
    \equiv \exp\!\left(-\oF(z)\right) 
\label{INT}
\end{equation}
or
\begin{equation}
    \oF(z) = (N+1)\ln z + \sum_{\nu=0}^{\infty}
    \ln\!\left(1 - ze^{-\beta\varepsilon_{\nu}}\right) \; ,
\label{LOG}
\end{equation}
the saddle-point $z_0$ is determined by the requirement that this function
becomes stationary,
\begin{equation}
    \left. \frac{\partial \oF(z)}{\partial z} \right|_{z=z_0} = 0 \; ,
\end{equation}
giving
\begin{equation}
    N + 1 = \sum_{\nu = 0}^{\infty}
    \frac{1}{z_0^{-1}e^{\beta\varepsilon_\nu} - 1} \; .
\label{SA1}
\end{equation}
Apart from the appearance of one extra particle on the left hand side,
this is just the grand canonical relation between particle number~$N$
and fugacity~$z_0$.

Proceeding according to folk wisdom, one then expands the logarithm $\oF(z)$
quadratically around $z_0$ and leads the path of integration parallel to the
imaginary axis over the saddle, relying on the fact that for large~$N$ the
main contribution to the integral~(\ref{ZIN}) is collected in the immediate
vicinity of the saddle-point~\cite{Schrodinger46}, so that the quadratic
expansion should prove sufficient. Doing the remaining Gaussian integral,
one arrives at the standard saddle-point approximation $\tZsp$ to the
canonical partition functions,       
\begin{eqnarray}
    \tZsp(\beta) & = &
    \frac{1}{2\pi i}\int_{z_0-i\infty}^{z_0+i\infty} \! {\rm d}z \,
    \exp\!\left(-\oF^{(0)} - \frac{1}{2}\oF^{(2)}(z-z_0)^2\right)
\nonumber \\
    & = & \frac{1}{2\pi}
    \exp(-\oF^{(0)}) \int_{-\infty}^{+\infty} \! {\rm d}u \, 
    \exp\!\left(+\frac{1}{2}\oF^{(2)}u^2\right)
\nonumber \\
    & = & \frac{\exp(-\oF^{(0)})}{\sqrt{-2\pi\oF^{(2)}}} \; .
\label{WSA}
\end{eqnarray}
Here and in the following we write $f^{(n)}$ for the $n$-th derivative
of a function~$f$ at a saddle-point; we have used $\oF^{(2)} < 0$. Hence,
one finds
\begin{eqnarray}
    \ln\tZsp(\beta) & = & -\frac{1}{2}\ln 2\pi - (N+1)\ln z_0
    - \sum_{\nu=0}^{\infty}
      \ln\!\left(1 - z_0e^{-\beta\varepsilon_{\nu}}\right)
\nonumber \\ & & 
    - \frac{1}{2} \ln \sum_{\nu=0}^{\infty}
      \frac{z_0^{-1} e^{-\beta\varepsilon_{\nu}}}
           {\left(1 - z_0e^{-\beta\varepsilon_{\nu}}\right)^2} \; ,
\label{LSP}
\end{eqnarray}
from which the canonical occupation number $\nacn$ of the state~$\alpha$
is obtained by differentiating once with respect to
$(-\beta\varepsilon_\alpha)$,
\begin{equation}
    \nacn = \frac{\partial\ln\tZsp}{\partial(-\beta\varepsilon_{\alpha})}
    + \frac{\partial\ln\tZsp}{\partial z_0}
      \frac{\partial z_0}{\partial(-\beta\varepsilon_{\alpha})} \; .
\end{equation}
As long as the gas is not condensed, the fourth term in the
approximation~(\ref{LSP}) remains small in comparison to the third,
and therefore may be neglected. Then the partial derivative  
$\partial\ln\tZsp/\partial z_0$ vanishes as a consequence of the
saddle-point equation~(\ref{SA1}), and one is left with
\begin{eqnarray}
    \nacn & = & \frac{\partial\ln\tZsp}{\partial(-\beta\varepsilon_{\alpha})}
\nonumber \\
    & = & \frac{1}{z_0^{-1}e^{\beta\varepsilon_{\alpha}}-1} \; ,
\label{WON}
\end{eqnarray}
so that within these approximations the canonical occupation numbers equal
their grand canonical counterparts.

However, this reasoning breaks down for temperatures below the onset
of Bose--Einstein condensation, where Eq.~(\ref{SA1}) requires that
$z_0^{-1}e^{\beta\varepsilon_0} - 1$ be on the order of $1/N$, so that 
the third and the fourth term on the right hand side of Eq.~(\ref{LSP})
are of comparable magnitude, namely of the order $O(\ln N)$. Hence, the
argument that led to the familiar formula~(\ref{WON}) becomes invalid. 
Moreover, inspecting the higher derivatives of $\oF(z)$ at the saddle-point,
\begin{equation}
    \oF^{(n)} = -\frac{(n-1)!}{z_0^n}\left[ (-1)^n(N+1)
    + \sum_{\nu=0}^{\infty}
    \left(\frac{1}{z_0^{-1}e^{\beta\varepsilon_{\nu}}-1}\right)^n\,\right] \; , 
\label{DER}
\end{equation}
one finds that in the condensate regime these derivatives grow dramatically
with increasing~$n$, $\oF^{(n)} = O(N^n)$, casting doubt on the validity
of the saddle-point approximation even if higher-order terms are
included~\cite{Schubert46}. However, for the derivation of an asymptotic
series the convergence properties of the formal Taylor series of $\oF(z)$
are irrelevant~\cite{Dingle73}; what actually endangers the approximation
scheme~(\ref{WSA}) in the condensate regime is the narrow approach of the
saddle-point to the ground-state singularity $z = e^{\beta\varepsilon_0}$
of the grand canonical partition function~(\ref{GPF}). Namely, the Gaussian
integral~(\ref{WSA}) can well represent the exact expression~(\ref{ZIN}) only
if the function $\oF(z)$ is free of singularities at least in those intervals
where the Gaussian is still large, i.e., where $-\oF^{(2)}(z - z_0)^2/2$ is
on the order of unity. Since $\oF^{(2)} = O(N^2)$, this observation translates
into the requirement that {\em the function $\oF(z)$ should be regular at
least in an interval of order $O(1/N)$ around $z_0$\/}. On the other hand,
{\em in the condensate regime the singularity at $z = e^{\beta\varepsilon_0}$
falls within a distance of order $O(1/N)$ from the saddle-point $z_0$\/},
again as a consequence of Eq.~(\ref{SA1}). It is this conflict, {\em not\/}
the poor behavior of the Taylor expansion of $\oF(z)$, which necessitates an
approach to the contour integral~(\ref{ZIN}) that is essentially more careful
than the standard scheme~(\ref{WSA}). Interestingly, though, the dilemma does
not appear to be overly severe --- the magnitude of both conflicting intervals
being of the {\em same\/} order $O(1/N)$ ---, and one might wonder already at
this point whether the usual procedure can be saved by simple means.         

The failure of the saddle-point approximation~(\ref{WSA}) for
condensed Bose gases has been the subject of a long debate in the
literature~\cite{Schubert46,FujiwaraEtAl70,ZiffEtAl77}, with notable early
contributions by Dingle~\cite{Dingle49} and Fraser~\cite{Fraser51}. Various
schemes have been designed for computing the number of condensate particles,
and its fluctuation, for gases with a fixed number~$N$ of particles,
{\em with\-out\/} resorting to the saddle-point method. Quite recently,
Navez {\em et al.\/} have suggested a statistical ensemble within which
one regards the condensate as an infinite particle reservoir for the
excited-states subsystem~\cite{NavezEtAl97}, thus putting into shape an idea
already expressed by Fierz~\cite{Fierz56}. In a mathematical setting, each
excited single-particle level $\varepsilon_\alpha$ then becomes formally
equivalent to a harmonic oscillator with frequency
$(\varepsilon_\alpha-\varepsilon_0)/\hbar$, which allows one to derive
elegant integral representations for canonical and microcanonical expectation
values: Setting $\varepsilon_0 = 0$ for convenience, restricting oneself to
temperatures below the onset of condensation, and introducing the spectral
Zeta function
\begin{equation}
    Z(\beta,t) = 
    \sum_{\nu=1}^{\infty} \frac{1}{(\beta\varepsilon_\nu)^t} \; ,
\label{SZF}
\end{equation}
with the sum running over the excited states only, the canonical number of
condensate particles can be written as~\cite{HKK98}
\begin{equation}
    \nzcn = N - \frac{1}{2\pi i} \int_{\tau-i\infty}^{\tau+i\infty} \!
    {\mbox d} t \, \Gamma(t) Z(\beta,t) \zeta(t)   \; ,
\label{ICN}
\end{equation}
where $\zeta(t)$ is the Riemann Zeta function. The path of integration
up the complex $t$-plane lies to the right of all poles of the integrand,
so that the residues of these poles, taken from right to left, yield a
systematic expansion of the integral~\cite{HKK98}. Similarly, the
canonical mean-square fluctuation of the number of condensate particles
takes the form    
\begin{equation}  
    \fzcn = \frac{1}{2\pi i}\int_{\tau-i\infty}^{\tau+i\infty} \!
    {\mbox d} t \, \Gamma(t) Z(\beta,t) \zeta(t-1)   \; ,
\label{ICF}
\end{equation}
and the difference between canonical and microcanonical fluctuations
is given by
\begin{equation}
    \fzcn - \fzmc =
    \frac{\left[ \frac{1}{2\pi i} \int_{\tau-i\infty}^{\tau+i\infty} \!
    {\mbox d} t \, \Gamma(t) Z(\beta,t-1) \zeta(t-1) \right]^2}
    {\frac{1}{2\pi i} \int_{\tau-i\infty}^{\tau+i\infty} \! {\mbox d} t \,
    \Gamma(t) Z(\beta,t-2) \zeta(t-1)} \; .
\label{IDF}
\end{equation}
Yet, there are at least two reasons not to be content with this state
of affairs. Firstly, since Eqs.~(\ref{ICN}), (\ref{ICF}), and (\ref{IDF})
rely on the presence of a reservoir of condensate particles, they are
blind to the onset of Bose--Einstein condensation, that is, to the sudden
appearance of this reservoir; as a consequence of this underlying ``oscillator
approximation'', they do not allow one to discuss just how such a sharp
feature can emerge in a Bose gas with a large, fixed number of particles.
Secondly, for a given single-particle spectrum the  integrals may not
always be straightforward to evaluate; already the treatment of an anisotropic
harmonic oscillator potential, in which case $Z(\beta,t)$ is related to Zeta
functions of the Barnes type~\cite{HKK98}, requires quite some analytical
skills. Therefore, one desires a tool that works, in principle, for all
temperatures, and is easy to use in practical calculations. 

The development of such a tool is the objective of the present paper. 
Following a trail pioneered by Dingle~\cite{Dingle73}, we work out and test
a variant of the saddle-point method that fulfills the above two requirements.  
Our guiding maxim is the same which already governed London's classic
analysis~\cite{London38} of the condensation phenomenon: If the ground state
is causing trouble, single it out and give it a special treatment --- which,
in our case, means to exempt the ground-state factor of the grand canonical
partition function~(\ref{GPF}) from the quadratic expansion performed in
the usual scheme~(\ref{WSA}), and to treat that factor exactly. As will be
demonstrated in detail, this natural strategy leads to an approach to the
canonical and the microcanonical statistics of condensed Bose gases which
is both extremely accurate and unsurpassed in computational ease.

We proceed as follows: In the next section we will first explain {\em why\/}
and {\em how\/} this proper saddle-point approximation works for the
calculation of the canonical partition functions, and how known limiting cases
are recovered, concentrating key technical details in Appendices~\ref{AA}
and~\ref{AB}. We then apply the method to computing canonical occupation
numbers and their fluctuations, illustrating the analytical results by
numerical calculations for an ideal Bose gas confined by an isotropic
harmonic potential. In Section~\ref{S3} we turn to the microcanonical
ensemble, and show how the very same refined saddle-point approach allows one
to obtain the interesting quantities almost without further effort. The final
Section~\ref{S4} summarizes the most important findings. 

Although our discussion is led along the lines of the ideal Bose gas, and
although some interesting physical insights will turn up --- in particular,
it will become clear why the canonical occupation numbers are well described
by the expression~(\ref{WON}), and thus essentially equal to their grand
canonical analogues, even for temperatures below the onset of condensation,
where the previous argument had failed ---, this is not primarily a work on
the ideal Bose gas as such. Rather, major emphasis lies on the mathematics
of saddle-point integrals with a singular integrand, of which condensed Bose
gases provide perhaps the most prominent examples; we hope that the detailed
exposition presented here will prove fruitful also in other areas of
mathematical physics where similar problems arise.

\section{The canonical ensemble}
\label{S2}

\subsection{The canonical partition function}

Since the conventional approximation~(\ref{WSA}) is thwarted by the fact
that the saddle-point~$z_0$, i.e., the solution to Eq.~(\ref{SA1}),
approaches the ground-state singularity $z = e^{\beta\varepsilon_0}$ of the
integrand~(\ref{INT}) within order $O(1/N)$ in the condensate regime, we now
exclude the ground-state contribution from the quadratic expansion of the
logarithm~(\ref{LOG}). That is, we define 
\begin{eqnarray}
    F(z) & = & \oF(z) - \ln\!\left(1 - ze^{-\beta\varepsilon_0}\right)
\nonumber \\
    & = & (N+1)\ln z + \sum_{\nu=1}^{\infty}
    \ln\!\left(1 - ze^{-\beta\varepsilon_{\nu}}\right) \; ,
\label{TEF}
\end{eqnarray}
and write the canonical $N$-particle partition function as
\begin{equation}
    Z_N(\beta) = \frac{1}{2\pi i} \oint \! {\rm d}z \,
    \frac{\exp\!\left(-F(z)\right)}{1 - ze^{-\beta\varepsilon_0}} \; .
\label{CPF}
\end{equation}
The key idea for treating integrals of this kind, due to
Dingle~\cite{Dingle73}, is to let the potentially dangerous denominator
stand as it is, and to expand only the tempered function $F(z)$ around the
saddle-point~$z_0$. The resulting approximation to $Z_N(\beta)$ will then
be valid for {\em all\/} temperatures: For high~$T$, when the saddle-point
moves away from the singularity, it is of no concern whether or not the
denominator is included in the Gaussian approximation; in the condensate
regime its exclusion is crucial. 

Let us first check the behavior of the derivatives $F^{(n)}$. When omitting
the ground-state term ($\nu = 0$) from the sum in Eq.~(\ref{DER}), the
behavior of the remaining sum is governed in the condensate regime by
the following terms ($\nu = 1,2,3,\ldots$), each of them being about
proportional to the $n$-th power of temperature. Hence, 	
\begin{equation}
    \left[ \sum_{\nu=1}^{\infty} \left(
    \frac{1}{z_0^{-1}e^{\beta\varepsilon_{\nu}}-1}\right)^n\,\right]^{1/n}
    \propto k_B T \; ,
\label{PR1}
\end{equation}
with approximate $T$-proportionality holding the better, the larger~$n$,
since large~$n$ emphasize the low-lying states. Now we focus on systems
with single-particle energies of the form~\cite{deGrootEtAl50}
\begin{equation}
    \varepsilon_{\{\nu_i\}} = \varepsilon_0
    + \varepsilon_1 \sum_{i=1}^{d} c_i \nu_i^s \; ,
\label{GRO}
\end{equation}
where $\nu_i = 0,1,2,\ldots$ are integer quantum numbers. The dimensionless
anisotropy coefficients $c_i$ should be of comparable magnitude, the lowest of
them equaling unity. This class of systems contains, e.g., a gas of $N$~ideal
Bose particles confined by a $d$-dimensional harmonic oscillator potential
($s = 1$), or by a $d$-dimensional hard box ($s = 2$). For $d/s > 1$ and
large~$N$ there is a sharp onset of Bose--Einstein condensation, with
\begin{equation}
    k_B T \propto \Nex^{s/d}
\label{PR2} 
\end{equation}
in the condensate regime; $\Nex \equiv N - \nzcn$ is the total number of
excited particles~\cite{GH97b,WeissWilkens97}. Thus, for largish~$n$
the Eqs.~(\ref{PR1}) and~(\ref{PR2}) give
\begin{equation}
    S_n \equiv
    \sum_{\nu=1}^{\infty} \left(
    \frac{1}{z_0^{-1}e^{\beta\varepsilon_{\nu}}-1}\right)^n
    \propto \Nex^{ns/d}
    \quad \mbox{for} \quad d/s > 1 \; ,
\label{SUM}
\end{equation}
resulting, by virtue of Eq.~(\ref{DER}), in
\begin{equation}
    F^{(n)} = O(N^{\xi(n)})
    \quad \mbox{with} \quad \xi(n) = \max\{1,ns/d\} \; ,
\label{ORD}
\end{equation}
as long as, besides $\nzcn = O(N)$, also $\Nex = O(N)$. To confirm this
relation, the case $n = 2$ of which will be of particular interest later,
Fig.~\ref{F_1} shows the numerically computed quantities
\begin{equation}
    r_n \equiv  \frac{\ln S_n}{n \, \ln \Nex}
\label{EXP}
\end{equation}
(which, in the general case, should approach $s/d$ for large~$n$) for a gas
of $N = 10^6$ ideal Bose particles kept at temperature $T = 0.5 \, T_0^{(3)}$
in a three-dimensional isotropic harmonic oscillator potential, with
\begin{equation}
    T_0^{(3)} = \frac{\hbar\omega}{k_B}\left(\frac{N}{\zeta(3)}\right)^{1/3}
\label{TT3}
\end{equation}
denoting the condensation temperature in the
large-$N$-limit~\cite{deGrootEtAl50}; $\omega$ is the oscillator frequency.
As expected, $r_n$ approaches the value $1/3$ fairly rapidly with~$n$. For
comparison, Fig.~1 also shows the corresponding data for a gas with the same
number of particles which is stored in a one-dimensional harmonic potential.
Its temperature is $T = 0.5 \, T_0^{(1)}$, where
\begin{equation}
    T_0^{(1)} = \frac{\hbar\omega}{k_B} \frac{N}{\ln N} \; .
\label{TT1}
\end{equation} 
This is a borderline case: For $d = s = 1$ there is no sharp onset of
Bose--Einstein condensation, so that $T_0^{(1)}$ merely plays the role
of a characteristic temperature below which the ground-state population
becomes significant. There are logarithmic corrections~\cite{HKK98} which
keep the ratios~$r_n$ below the value $s/d = 1$ also for large $n$, but,
as seen in the figure, even now $r_n$ rapidly approaches a constant not too
far from unity. We conclude that for systems of the type~(\ref{GRO}),
with $d/s > 1$, the $O(N^n)$-growth of $\oF^{(n)}$ in the condensate regime
is replaced by the somewhat milder $O(N^{ns/d})$-growth of $F^{(n)}$ when
going from $\oF(z)$ to its ground-state-amputated descendant~$F(z)$, so that
also the formal Taylor expansion of~$F(z)$ around the saddle-point~$z_0$
is ill-behaved.

But as already indicated, the properties of the Taylor series of $F(z)$ are
only of secondary importance. What really matters is that this function does
not share the ground-state singularity; the singular point to be watched now is
the one at $z = e^{\beta\varepsilon_1}$. Since $z_0 < e^{\beta\varepsilon_0}$,
the saddle-point remains separated from that singularity by at least the
$N$-independent gap $e^{\beta\varepsilon_1} - e^{\beta\varepsilon_0}
\approx (\varepsilon_1 - \varepsilon_0)/k_B T$. This guarantees that when
quadratically expanding the amputated function $F(z)$, rather than $\oF(z)$,
around $z_0$, an interval of the required order $O(1/\sqrt{-F^{(2)}}) =
O(N^{-\xi(2)/2})$ becomes singularity-free for sufficiently large~$N$;
the higher the temperature (while remaining in the condensate regime), the
smaller the gap, and the larger the particle number has to be. Then the
Gaussian approximation to $\exp(-F(z))$ is safe. As shown in detail in
Appendix~\ref{AA}, the subsequently emerging saddle-point integral for the
canonical partition function~(\ref{CPF}) can be done exactly, yielding
(cf.\ Eq.~(\ref{CSA}) with $\sigma = 1$)
\begin{equation}
    \Zsp(\beta) = \frac{1}{\sqrt{2\pi}}
    \exp\!\left(\beta\varepsilon_0 - F^{(0)} - 1
        + \frac{1}{2}\eta^2 - \frac{1}{4}\eb^2 \right) D_{-1}(\eb) \; ,
\label{CZN}
\end{equation}
where			
\begin{eqnarray}
    \eta & = & \left(e^{\beta\varepsilon_0} - z_0 \right)\sqrt{-F^{(2)}} \; ,
\label{CXN}
\\
    \eb & = & \eta - \frac{1}{\eta} \; ,
\label{CXW} 
\end{eqnarray}
and $D_{-1}(\eb)$ is a parabolic cylinder function, employing Whittaker's
notation~\cite{WhittakerWatson62,RyshikGradstein63}. For discussing this
unfamiliar-looking expression~(\ref{CZN}), which, as already remarked above,
is valid for {\em all\/} temperatures, we observe~\cite{RyshikGradstein63}
that $D_{-1}$ is related to the complementary error function ${\rm erfc}$,
\begin{equation}
    D_{-1}(\eb) = \exp\!\left(\frac{1}{4}\eb^2\right)
    \sqrt{\frac{\pi}{2}} \, {\rm erfc}\!\left(\frac{\eb}{\sqrt{2}}\right) \; ;
\end{equation}
hence
\begin{equation}
    \Zsp(\beta) = \exp\!\left(\beta\varepsilon_0 - F^{(0)} - 1
    + \frac{1}{2}\eta^2 \right)
    \frac{1}{2} {\rm erfc}\!\left(\frac{\eb}{\sqrt{2}}\right) \; .
\label{CEF}
\end{equation}
For high temperatures, well above the condensation point, $z_0$ approaches
zero, so that the parameter $\eta$, and as a consequence also $\eb$, grows
without bound when $N$ becomes large. Then we may replace the complementary
error function by the leading term of its asymptotic expansion for large
{\em positive\/} arguments~\cite{AbramowitzStegun72},
\begin{equation}
    {\rm erfc}\!\left(\frac{\eb}{\sqrt{2}}\right) \sim
    \sqrt{\frac{2}{\pi}} \frac{\exp(-\eb^2/2)}{\eb} \; .
\end{equation}
This is a special case of the approximation~(\ref{HTA}) introduced in
Appendix~\ref{AB}, and implies, together with the further approximations
(\ref{HTS}) -- (\ref{HTF}), that the general expression~(\ref{CZN})
correctly approaches the standard saddle-point result~(\ref{WSA}) outside
the condensate phase,   
\begin{equation}
    \Zsp(\beta) \sim \tZsp(\beta)
    \qquad \mbox{for high} \; T \; . 
\end{equation}
This was to be expected, since when the denominator in the
integrand~(\ref{CPF}) does not become small, it doesn't matter whether it
is included in the quadratic approximation, as in the scheme~(\ref{WSA}),
or treated exactly, as in the derivation of Eq.~(\ref{CZN}).

In the condensate regime, where $e^{\beta\varepsilon_0} - z_0 = O(1/N)$
and $F^{(2)} = O(N^{\xi(2)})$ as specified by Eq.~(\ref{ORD}), the
definition~(\ref{CXN}) of the parameter~$\eta$ implies either
$\eta = O(N^{-1/2})$ or $\eta = O(N^{s/d-1})$, whichever is larger. Since
we require $d/s > 1$, we conclude that in either case $\eta$~approaches
zero for large~$N$, so that now $\eb = \eta - \eta^{-1}$ is a large
{\em negative\/} number. Therefore, we may safely use the approximation
${\rm erfc}(\eb/\sqrt{2}) \approx 2$, and arrive at 
\begin{eqnarray}
    \Zsp(\beta) & \sim &
    \exp\!\left(\beta\varepsilon_0 - F^{(0)} - 1\right)
\nonumber \\ & = & 
    \frac{e^{\beta\varepsilon_0-1}}{z_0^{N+1}}
    \prod_{\nu=1}^{\infty} \frac{1}{1 - z_0\exp(-\beta\varepsilon_{\nu})} \; .
\label{COA}
\end{eqnarray}
Even when the particle number~$N$ is merely moderately large, the decrease
of the parameter~$\eb$ with temperature from large positive to large negative
values can be fairly rapid, so that the complementary error function in the
canonical $N$-particle partition function~(\ref{CEF}) acts as a switch,
meaning that the transition from the high-temperature asymptotics~(\ref{WSA})
to the condensate asymptotics~(\ref{COA}) becomes quite sharp. This is
confirmed by Fig.~\ref{F_2}, which depicts ${\rm erfc}(\eb/\sqrt{2})$ as
function of temperature for a gas of $N = 10^3$ ideal Bosons in a
three-dimensional isotropic harmonic potential. In the borderline case of
the one-dimensional harmonic potential, also indicated in the figure, there
is no such sharp transition.

It is worthwhile to discuss the condensate partition function~(\ref{COA})
a little further. Anticipating that, despite the incorrect reasoning, the
expression~(\ref{WON}) for the canonical occupation numbers will remain
valid approximately even in the condensate regime, we have	 
\begin{equation}
    z_0^{-1}e^{\beta\varepsilon_0} = 1 + \nzcn^{-1} \; .
\label{NZC}
\end{equation}
Hence, we may eliminate the saddle-point~$z_0$ by setting
\begin{eqnarray}
    \left(\frac{1}{z_0}\right)^{N+1} & = & e^{-(N+1)\beta\varepsilon_0}
    \left(1 + \nzcn^{-1}\right)^{N+1}
\nonumber \\
    & \sim & e^{-(N+1)\beta\varepsilon_0} e^{N/\nzcn} \; ,
\end{eqnarray}
yielding  
\begin{equation}
    \Zsp(\beta) \sim \exp\!\left(N/\nzcn - 1 - N\beta\varepsilon_0\right) \,
    \prod_{\nu=1}^{\infty}
    \frac{1}{1 - \exp(\beta[\varepsilon_0-\varepsilon_{\nu}])} \; .
\label{OPF}
\end{equation}
The infinite product on the right hand side, describing the excited-states
subsystem, equals the canonical partition function of a collection of
infinitely many, distinguishable harmonic oscillators with frequencies
$(\varepsilon_\nu - \varepsilon_0)/\hbar$ (where $\nu = 1,2,3,\ldots$),
thus leading back to the ``oscillator approximation'' which has been the
starting point for the derivation of the integral representations~(\ref{ICN}),
(\ref{ICF}), and (\ref{IDF}) in Ref.~\cite{HKK98}. In contrast, the value
of the present approximation~(\ref{OPF}) lies in the fact that it is not
restricted to the excited states, but also contains the ground state
explicitly.  

Taking the derivatives with respect to $-\beta\varepsilon_\alpha$, with
$\alpha \neq 0$, one then finds occupation numbers 
\begin{eqnarray}
    \nacn & = & \frac{\partial\ln\Zsp}{\partial(-\beta\varepsilon_{\alpha})}
\nonumber \\
     & = & \frac{1}{e^{\beta(\varepsilon_{\alpha}-\varepsilon_0)}-1}
\label{OAN}
\end{eqnarray}
and mean-square fluctuations	
\begin{eqnarray}
    \facn & = & \frac{\partial^2\ln\Zsp}
                     {\partial(-\beta\varepsilon_{\alpha})^2}
\nonumber \\
          & = & \nacn \left( \nacn + 1 \right)
\label{OAF}
\end{eqnarray}
of the excited states, while differentiating the logarithm of Eq.~(\ref{OPF})
with respect to $-\beta\varepsilon_0$ produces first the obvious identity
\begin{equation}
    \nzcn = N - \sum_{\nu=1}^{\infty} \, \nncn
\label{OAG}
\end{equation}
and then the important equation
\begin{equation}
    \fzcn = \sum_{\nu=1}^{\infty} \, \fncn \; ;
\end{equation}
stating that, in the condensate regime and subject to the above
approximations, within the canonical ensemble the occupation numbers
of the excited states are {\em uncorrelated\/} stochastic
variables~\cite{Dingle49,Fraser51,HKK98}. 

Remarkably, the error one would have committed had one naively employed
the standard approximation~(\ref{WSA}) in the condensate regime, and
which can be quantified only now, is not devastating. As explained
in Appendix~\ref{AB} (cf.~Eq.~(\ref{LTE})), {\em in the low-temperature,
large-$N$-regime the incorrect partition function $\tZsp(\beta)$ exceeds
the correct partition function by merely the temperature-independent factor
$1/R_1 \approx 1.08444$, regardless of the single-particle spectrum\/}, that
is, regardless of the trapping potential --- implying that the error might
even go unnoticed when carelessly taking derivatives of $\ln \tZsp(\beta)$.
This finding is illustrated in Fig.~\ref{F_3}, again for an ideal Bose gas
with $N = 1000$ particles in a three-dimensional isotropic harmonic potential.
The figure shows the ratios of $\tZsp(\beta)$, computed according to the
scheme~(\ref{WSA}), of the correct approximation~(\ref{CZN}), and of its
condensate descendant~(\ref{COA}), to the exact canonical partition function,
the latter having been obtained from the one-particle partition function
$Z_1(\beta)$ by means of the familiar recursion
formula~\cite{WeissWilkens97,Landsberg61,BF93,BDL96,BalazsBergeman98}
\begin{equation}
    Z_N(\beta) = \frac{1}{N}\sum_{k=1}^{N} Z_1(k\beta) Z_{N-k}(\beta)
    \quad , \quad Z_0(\beta) \equiv 1 \; .
\label{FRF}    
\end{equation}
As witnessed by Fig.~\ref{F_3}, the quality of the proper
approximation~(\ref{CZN}) is outstanding for all temperatures, its
low-temperature variant~(\ref{COA}) performs bravely where it is expected to,
and the standard approximation~(\ref{WSA}) is good at high temperatures, but
fails by the predicted factor $1/R_1$ in the condensate regime. For comparison,
Fig.~\ref{F_4} shows the corresponding data for the one-dimensional case;
here the exact $N$-particle partition function is known in closed
form~\cite{AuluckKothari46,TodaEtAl92}. We find features that are
qualitatively similar to those in the preceding figure, although, with
$N = 1000$, the approximation~(\ref{CZN}) is not quite as good at
intermediate~$T$. The standard scheme~(\ref{WSA}) again is off by the same,
universal factor $1/R_1$ at low temperatures.

\subsection{Canonical occupation numbers}

The derivation of the expressions~(\ref{OAN}) and~(\ref{OAG}) from the
condensate approximation~(\ref{COA}) to the canonical partition function
serves to render that partition function plausible, but it is not the best
one can do, since we have simply {\em assumed\/} the validity of
Eq.~(\ref{NZC}). An accurate and fully consistent computation of these
occupation numbers starts from the identity 
\begin{eqnarray}
   \nacn & = & \frac{\partial}{\partial(-\beta\varepsilon_{\alpha})}
   \ln Z_N(\beta)
\nonumber \\ & = &
   \frac{1}{Z_N(\beta)} \, \frac{1}{2\pi i} \oint \! {\rm d}z \,
   \frac{1}{z^N} \prod_{\nu=0}^{\infty}
   \frac{1}{1 - z\exp(-\beta\varepsilon_{\nu})} \,
   \frac{\exp(-\beta\varepsilon_{\alpha})}
        {1 - z\exp(-\beta\varepsilon_{\alpha})}
\nonumber \\ & \equiv &
    \frac{1}{Z_N(\beta)} \, \frac{1}{2\pi i} \oint \! {\rm d}z \,
    \exp\!\left(-\oG(z)\right) \; ,
\label{CON}
\end{eqnarray}
with
\begin{equation}
    \oG(z) = N\ln z
    + \sum_{\nu=0}^{\infty}\ln\!\left(1 - ze^{-\beta\varepsilon_{\nu}}\right)
    + \ln\!\left(1 - ze^{-\beta\varepsilon_{\alpha}}\right)
    + \beta\varepsilon_{\alpha} \; .
\end{equation}
The equation that determines the saddle-point~$z_1$ for the new contour
integral~(\ref{CON}), namely	
\begin{equation}
    N = \sum_{\nu = 0}^{\infty}
      \frac{1}{z_1^{-1}e^{\beta\varepsilon_\nu} - 1}
    + \frac{1}{z_1^{-1}e^{\beta\varepsilon_\alpha} - 1} \; ,
\label{NSP}
\end{equation}
formally looks like the grand canonical relation between particle
number~$N$ and fugacity $z_1$ for a system with an extra energy level
$\varepsilon_\alpha$. Now we have to distinguish two cases:

If $\alpha \neq 0$, we merely have to copy the steps made in the derivation
of the proper canonical partition function~(\ref{CZN}). That is, we separate
the ground-state contribution from the exponent $\oG(z)$ by defining the
tempered function
\begin{equation}
    G(z) = \oG(z) - \ln\!\left(1 - ze^{-\beta\varepsilon_0}\right) \; ,
\label{TEG}
\end{equation}
and obtain    
\begin{eqnarray}
    \frac{\partial}{\partial(-\beta\varepsilon_{\alpha})} Z_N(\beta)
    & = & \frac{1}{2\pi i} \oint \! {\rm d}z \,
    \frac{\exp\!\left(-G(z)\right)}{1 - ze^{-\beta\varepsilon_0}}
\nonumber \\
    & \sim & \exp\!\left(\beta\varepsilon_0 - G^{(0)} - 1\right) \; ,
\label{DZC}
\end{eqnarray}
proceeding at once to temperatures below the onset of condensation.
Then Eq.~(\ref{CON}), together with the previous result~(\ref{COA}) for
$Z_N(\beta)$, yields the expression
\begin{equation}
    \nacn = \exp\!\left(F(z_0) - G(z_1)\right)	\qquad (\alpha \neq 0)
\label{CSN}
\end{equation}
for the canonical occupation numbers of the excited states in the
condensate regime. This accurate result is well approximated by the
previous formula~(\ref{OAN}): Since the level that has artificially
been doubled differs from the ground state, we may set
$z_0 \approx z_1 \approx e^{\beta\varepsilon_0}$ in Eq.~(\ref{CSN}),
and obtain 
\begin{eqnarray}
    \nacn & \approx & \exp\!\left(\ln z_0 - \ln(1 - z_0
    e^{-\beta\varepsilon_{\alpha}}) - \beta\varepsilon_{\alpha}\right)
\nonumber \\
    & \approx & \frac{1}{e^{\beta(\varepsilon_{\alpha}-\varepsilon_0)}-1}
    \; ,
\label{ZEZ}
\end{eqnarray}	
using the definitions~(\ref{TEF}) and (\ref{TEG}) of the functions~$F$
and~$G$.

If, however, $\alpha = 0$, the exponent $\oG(z)$ in Eq.~(\ref{CON})
corresponds to a system with a doubled ground state. Hence, we have to
temper this function accordingly, and define
\begin{equation}
    G(z) = \oG(z) - \ln\!\left(e^{\beta\varepsilon_0}
    [1 - ze^{-\beta\varepsilon_0}]^2\right) \; .
\end{equation}
This leads to
\begin{eqnarray}
    \frac{\partial}{\partial(-\beta\varepsilon_0)} Z_N(\beta) & = &
    \frac{1}{2\pi i} \oint \! {\rm d}z \,
    \frac{\exp\!\left(-G(z)-\beta\varepsilon_0\right)}
         {\left(1 - ze^{-\beta\varepsilon_0}\right)^2}
\nonumber \\
    & \sim & \frac{2}{e^{\beta\varepsilon_0} - z_1}
    \exp\!\left(\beta\varepsilon_0 - G^{(0)} - 2\right) \; , 
\end{eqnarray}
where we have employed the condensate approximation~(\ref{LTA}) to the
general saddle-point formula~(\ref{CSA}), with $\sigma=2$. Thus, in the
condensate regime the canonical ground-state occupation number is given by   
\begin{equation}
    \nzcn = \frac{2}{z_1^{-1}e^{\beta\varepsilon_0} - 1}
    \exp\!\left(F(z_0) - G(z_1) - 1 - \ln z_1\right) \; .
\label{CSZ}
\end{equation}
For showing that this cumbersome expression actually is consistent with
the familiar grand-canonical result in the large-$N$-limit, we now
have to carefully keep track of the two different parameters $z_0$ and
$z_1$. That is, we may set
\begin{equation}
    \frac{1}{z_0^{-1}e^{\beta\varepsilon_0} - 1} 
    \approx \nzcn \; ,
\label{PL1}
\end{equation}
whereas
\begin{equation}
    \frac{1}{z_1^{-1}e^{\beta\varepsilon_0} - 1}
    \approx \frac{\nzcn}{2} \; ,
\label{PL2}
\end{equation}
reflecting the double appearance of the ground state in Eq.~(\ref{NSP}).
Hence, the argument of the exponential in the ground-state occupation
formula~(\ref{CSZ}) should be approximately equal to zero. This follows
by observing  
\begin{eqnarray}
    & & F(z_0) - G(z_1) - 1 - \ln z_1
\nonumber \\
    & = & (N+1)\ln\frac{z_0}{z_1} - 1 + \sum_{\nu=1}^{\infty} \ln\!\left(
      \frac{1 - z_0 e^{-\beta\varepsilon_{\nu}}}
           {1 - z_1 e^{-\beta\varepsilon_{\nu}}}\right) 
\nonumber \\
    & \approx & (N+1)\ln\frac{z_0}{z_1} - 1
    - \left(1 - \frac{z_1}{z_0}\right)
      \sum_{\nu=1}^{\infty} \frac{1}{z_0^{-1}e^{\beta\varepsilon_0} - 1}
\nonumber \\
    & = & (N+1)\ln\frac{z_0}{z_1} - 1
    - \left(1 - \frac{z_1}{z_0}\right) \left(N + 1 - \nzcn \right) \; ,  
\end{eqnarray}
which, upon inserting 
\begin{equation}
    \frac{z_0}{z_1} \approx 1 + \frac{1}{\nzcn} 
\end{equation}
as obtained from Eqs.~(\ref{PL1}) and~(\ref{PL2}), indeed gives the
required relation 
\begin{equation}
    F(z_0) - G(z_1) - 1 - \ln z_1 \approx 0 \; .
\end{equation}

For demonstrating the accuracy of the canonical formulas~(\ref{CSN})
and~(\ref{CSZ}), we resort once more to a gas of $N = 1000$ ideal Bose
particles in a three-dimensional harmonic oscillator potential.
Figure~\ref{F_5} shows the occupation number $\nocn$ as a function of
temperature, computed according to Eq.~(\ref{CSN}), and compares these
data to those that are obtained if $Z_N(\beta)$ and
$\partial Z_N(\beta)/\partial(-\beta\varepsilon_1)$ are naively calculated
from the standard saddle-point formula. The inset quantifies the ratios
of these approximate occupation numbers to the exact ones, which have
again been computed recursively. In the condensate regime, the naive
approximation to {\em both\/} $Z_N(\beta)$ and
$\partial Z_N(\beta)/\partial(-\beta\varepsilon_1)$ is off by the same
universal factor $1/R_1$ derived in Appendix~\ref{AB}, so that this error
cancels when forming their ratio: The standard saddle-point scheme
accidentally yields the correct canonical occupation numbers of the excited
states both above and below the onset of condensation. On the other hand,
if one uses even at high temperatures the approximations~(\ref{COA})
and~(\ref{DZC}), each of which is correct in the condensate regime only, it
follows from Eq.~(\ref{HTE}) that the ratio of the individual errors is given
by the square root of ${\oG^{(2)}/\oF^{(2)}}$. Since this ratio approaches
unity in the large-$N$, high-$T$-regime, Eq.~(\ref{CSN}) actually is correct
also at high temperatures.  

This accidental correctness is no longer met in the case of the ground-state
occupation number $\nzcn$, depicted in Fig.~6. Now the error introduced
when computing $\partial Z_N(\beta)/\partial(-\beta\varepsilon_0)$ at low
temperatures within the standard scheme is given by the factor $1/R_2$
(see Appendix~\ref{AB}), so that the resulting value of $\nzcn$ is too
small by the constant factor $R_1/R_2 \approx 0.96106$ in the condensate
regime, whereas Eq.~(\ref{CSZ}) yields the correct data. Outside the
condensate regime the standard approximation becomes correct, whereas
Eq.~(\ref{HTE}) reveals that the condensate approximation~(\ref{CSZ}) is
wrong by the factor $M_2/M_1 \sim R_2/(R_1\sqrt{2}) \approx 0.73576$. 
Needless to say, if one deduces the canonical occupation numbers directly
from the ratios of the saddle-point approximations~(\ref{CSA}) to  
the respective contour integrals, instead of invoking their high- and
low-temperature limits~(\ref{USR}) and~(\ref{LTA}), one obtains expressions
that are valid for all temperatures.

\subsection{Canonical fluctuations}

The calculation of the canonical mean-square fluctuations $\facn$ of
the occupation numbers now directly parallels that of the occupation
numbers themselves: Starting from the identity 
\begin{eqnarray}
    \facn & = & \frac{\partial^2\ln Z_N}
                     {\partial(-\beta\varepsilon_{\alpha})^2}
\nonumber \\
    & = & \frac{1}{Z_N} \frac{\partial Z_N}
                             {\partial(-\beta\varepsilon_{\alpha})}
    - \left(\frac{1}{Z_N} \frac{\partial Z_N}
                             {\partial(-\beta\varepsilon_{\alpha})}\right)^2
    + \frac{2}{Z_N} \, \frac{1}{2\pi i} \oint \! {\rm d}z \,
      \exp\!\left(-\oH(z)\right) \; ,
\label{COF}
\end{eqnarray}
we are left with the task to evaluate the further contour integral
\begin{equation}
    \frac{1}{2\pi i} \oint \! {\rm d}z \, \exp\!\left(-\oH(z)\right) \equiv
    \frac{1}{2\pi i} \oint \! {\rm d}z \,
    \frac{1}{z^{N-1}} \prod_{\nu=0}^{\infty}
    \frac{1}{1 - z\exp(-\beta\varepsilon_{\nu})} \, 
    \frac{\exp(-2\beta\varepsilon_{\alpha})}
	 {\left[1 - z\exp(-\beta\varepsilon_{\alpha})\right]^2} \; ,
\end{equation}
where
\begin{equation}
    \oH(z) = (N-1)\ln z
    + \sum_{\nu=0}^{\infty}\ln\!\left(1 - ze^{-\beta\varepsilon_{\nu}}\right)
    + 2\ln\!\left(1 - ze^{-\beta\varepsilon_{\alpha}}\right)
    + 2\beta\varepsilon_{\alpha} \; .
\end{equation}
As a consequence of the second derivative performed in Eq.~(\ref{COF}),
the state with energy~$\varepsilon_\alpha$ now has formally been tripled,
giving the equation
\begin{equation}
    N-1 = \sum_{\nu = 0}^{\infty}
    \frac{1}{z_2^{-1}e^{\beta\varepsilon_\nu} - 1}
    + \frac{2}{z_2^{-1}e^{\beta\varepsilon_\alpha} - 1}
\label{SA3}
\end{equation}
for the new saddle-point $z_2$. Thus, for $\alpha \neq 0$ the tempered
version of the function $\oH(z)$ becomes 
\begin{equation}
    H(z) = \oH(z) - \ln\!\left(1 - ze^{-\beta\varepsilon_0}\right) \; ,
\end{equation}
implying
\begin{eqnarray}
    \frac{1}{2\pi i} \oint \! {\rm d}z \, \exp\!\left(-\oH(z)\right) & = &
    \frac{1}{2\pi i} \oint \! {\rm d}z \,
    \frac{\exp\!\left(-H(z)\right)}{1 - ze^{-\beta\varepsilon_0}}
\nonumber \\
    & \sim & \exp\!\left(\beta\varepsilon_0 - H^{(0)} - 1\right)
\end{eqnarray}
in the condensate regime; hence
\begin{equation}
   \facn = \nacn - \nacn^2 + 2\exp\!\left(F(z_0) - H(z_2)\right)
   \qquad (\alpha \neq 0) \; .
\label{CFN}
\end{equation}
Setting $z_0 \approx z_2 \approx e^{\beta\varepsilon_0}$, similar to
the reasoning underlying Eq.~(\ref{ZEZ}), it can be seen that the third
term on the right hand side is close to $2\nacn^2$, so that this
expression~(\ref{CFN}) properly reduces to the familiar Eq.~(\ref{OAF}).  

If $\alpha  = 0$, we disentangle the three-fold ground-state contribution
from $\oH(z)$ by defining
\begin{equation}
    H(z) = \oH(z) - \ln\!\left(e^{2\beta\varepsilon_0}
    [1 - ze^{-\beta\varepsilon_0}]^3\right) \; ,
\end{equation}
and get	
\begin{eqnarray}
    \frac{1}{2\pi i} \oint \! {\rm d}z \, \exp\!\left(-\oH(z)\right) & = &
    \frac{1}{2\pi i} \oint \! {\rm d}z \,
    \frac{\exp\!\left(-H(z)-2\beta\varepsilon_0\right)}
         {\left(1 - ze^{-\beta\varepsilon_0}\right)^3}
\nonumber \\
    & \sim &
    \frac{1}{2}\left(\frac{3}{e^{\beta\varepsilon_0}-z_2}\right)^2
    \exp\!\left(\beta\varepsilon_0 - H^{(0)} - 3\right) \; ,
\end{eqnarray}
employing the condensate approximation~(\ref{LTA}) with $\sigma = 3$.
Therefore, the canonical fluctuation of the number of condensate particles
now takes the form 
\begin{equation}
    \fzcn = \nzcn - \nzcn^2 +
    \frac{9}{\left(e^{\beta\varepsilon_0}-z_2\right)^2}
    \exp\!\left(F(z_0) - H(z_2) - 2\right) \; .
\label{CFZ}
\end{equation}

We check the results~(\ref{CFN}) and~(\ref{CFZ}) again for $N = 1000$ ideal
Bosons in an isotropic harmonic oscillator potential. Figure~\ref{F_7} shows
the root-mean-square fluctuation $\mocn \equiv \focn^{1/2}$ as obtained
from Eq.~(\ref{CFN}), and from the standard saddle-point approximations to
the three individual terms on the right hand side of the identity~(\ref{COF}).
For each term we have the same accidental correctness of the standard
saddle-point result in the condensate regime, and of the condensate
approximation at high temperatures, as already described for the occupation
numbers of the excited states; the inset, which depicts the ratios of the two
approximations to the recursively calculated exact fluctuation, confirms that
either approximation is correct at both high and low temperatures.   
 
In the case of the condensate fluctuation, however, the situation is
quite different. As witnessed by the inset in Fig.~\ref{F_8}, our
formula~(\ref{CFZ}) gives the correct fluctuation of the number of
ground-state particles in the condensate regime. Since the canonical
mean-square fluctuation $\fzcn$ has to vanish for zero temperature, when
all $N$~particles occupy the ground state, the third term on the right hand
side of Eq.~(\ref{CFZ}) approaches $N^2 - N$ for $T \to 0$. When naively
using the standard saddle-point scheme, the results for the individual terms
in Eq.~(\ref{COF}) are incorrect by factors $R_1/R_2$, $(R_1/R_2)^2$, and
$R_1/R_3$, respectively, with the universal numbers $R_\sigma$ determined
in Appendix~\ref{AB}. Hence, the naive scheme yields spurious mean-square
ground-state fluctuations
\begin{eqnarray}
    \fzcn^{\rm{spur}} & = &
    [R_1/R_3 - (R_1/R_2)^2]N^2 + [R_1/R_2 - R_1/R_3]N
\nonumber \\
    & \approx & 0.02438 \, N^2 + 0.01304 \, N
\label{ERR}
\end{eqnarray}
for $T \to 0$, giving, for instance, the incorrect r.m.s.-fluctuation
$\mzcn^{\rm{spur}} \approx 156$ for $N = 1000$, in precise agreement with
what is observed in Fig.~\ref{F_8}. In the opposite regime, that is, for
high temperatures, the standard scheme becomes correct. Then, since
$\nzcn \ll 1$, the condensate approximation~(\ref{CFZ}) is off by roughly
the same factor $R_2/(R_1\sqrt{2}) \approx 0.73576$ that also determines
the error of $\nzcn$ itself.   

It is now also illuminating to compare the saddle-point scheme developed
here to the oscillator approximation that has led in Ref.~\cite{HKK98}
to the integral representations~(\ref{ICN}) -- (\ref{IDF}). This latter
approximation cannot cope with the Bose--Einstein transition, that is, its
validity remains restricted to the condensate regime, since it rests on the
fiction of an infinite reservoir of condensate particles~\cite{NavezEtAl97}.
This very feature, however, is what allows one to derive closed expressions
for the condensate fluctuations, provided the pole structure of the spectral
Zeta function~(\ref{SZF}) in the complex $t$-plane is known. For an ideal
Bose gas in a three-dimensional isotropic harmonic potential, this function
can be written in terms of Riemann Zeta functions,
\begin{equation}
    Z(\beta,t) = (\beta\hbar\omega)^{-t} \left[ \frac{1}{2}\,\zeta(t-2)
    + \frac{3}{2}\,\zeta(t-1) + \zeta(t) \right] \; ,
\label{OZF}
\end{equation}
having shifted the ground-state energy to $\varepsilon_0 = 0$. Taking into
account the three rightmost poles of the product
$\Gamma(t) Z(\beta,t) \zeta(t-1)$, located at $t=3$, $t=2$, and $t=1$,
Eq.~(\ref{ICF}) then gives  
\begin{eqnarray}
    \fzcn & = & \left(\frac{k_BT}{\hbar\omega}\right)^3 \zeta(2)
\nonumber \\
    & + & \left(\frac{k_BT}{\hbar\omega}\right)^2
    \left[\frac{3}{2}\ln\!\left(\frac{k_BT}{\hbar\omega}\right)
          + \frac{3}{2}\gamma + \frac{5}{4} + \zeta(2) \right]
\nonumber \\
    & + & \;\; \frac{k_BT}{\hbar\omega} \left(-\frac{1}{2}\right) 
\label{OAP}
\end{eqnarray}
for $k_BT/(\hbar\omega) \gg 1$, where $\gamma \approx 0.57722$ is Euler's
constant. In contrast, the saddle-point approach requires the numerical
determination of the three saddle-point parameters $z_0$, $z_1$, and $z_2$
from Eqs.~(\ref{SA1}), (\ref{NSP}), and (\ref{SA3}), respectively, but this
effort is rewarded by the possibility to monitor the fluctuations for all
temperatures, including the transition regime. As detailed in Fig.~\ref{F_9},
sufficiently below the onset of condensation both the saddle-point
approximation~(\ref{CFZ}) and the oscillator approximation~(\ref{OAP}) yield
excellent agreement with the exact condensate fluctuation, even for particle
numbers as low as $N = 1000$.

\section{The microcanonical ensemble}
\label{S3}

\subsection{The microcanonical partition function}

For extending the techniques developed in the previous section to
the microcanonical ensemble, we write the grand canonical partition
function~(\ref{GPF}) as
\begin{equation}
   \prod_{\nu=0}^{\infty} \frac{1}{1 - z\exp(-\beta\varepsilon_{\nu})} =
   \sum_{N=0}^{\infty} z^N \sum_{E} e^{-\beta E} \,
   \widetilde{\Omega}(E,N) \; ,
\label{GPM}
\end{equation}
where the microcanonical partition functions $\widetilde{\Omega}(E,N)$
denote the number of microstates accessible to a thermally isolated
$N$-particle system with total excitation energy~$E$. Setting the
ground-state energy equal to zero, we now assume that all single-particle
energies $\varepsilon_{\nu}$, and hence also all possible excitation energies,
can be represented reasonably well as integer multiples of a basic quantum
$\hbar\omega$. Introducing the variable
\begin{equation}
    x = e^{-\beta\hbar\omega}
\end{equation}
and writing $E/(\hbar\omega) = m$, Eq.~(\ref{GPM}) takes the form of a
double power series, 
\begin{eqnarray}
   \prod_{\nu=0}^{\infty} \frac{1}{\left(1 - z x^{\nu}\right)^{g_{\nu}}}
   & = & \sum_{N=0}^{\infty} z^N \sum_{m=0}^{\infty} x^m \, \Omega(m,N)
\nonumber \\
   & \equiv & \Xi(x,z) \; ,   
\label{DSE}
\end{eqnarray}
with $g_{\nu}$ indicating the degree of degeneracy of the single-particle
energy level $\nu \hbar\omega$, and
$\Omega(m,N) \equiv \widetilde{\Omega}(m\hbar\omega,N)$. With the help of two
suitable contours which encircle the origin of the complex $x$- and $z$-plane,
respectively, the desired partition functions can be isolated from this series
by means of the identity
\begin{eqnarray}
    \Omega(m,N) & = &
    \frac{1}{(2\pi i)^2} \oint \! {\rm d}x \oint \! {\rm d}z \,
    \frac{1}{x^{m+1}z^{N+1}} \prod_{\nu=0}^{\infty}
    \frac{1}{\left(1 - z x^{\nu}\right)^{g_{\nu}}}
\nonumber \\
    & \equiv & \frac{1}{(2\pi i)^2} \oint \! {\rm d}x \oint \! {\rm d}z \,
    \exp\!\left(-\oF(x,z)\right) \; ,
\label{DCI}
\end{eqnarray}
where
\begin{equation}
    \oF(x,z) = (m+1)\ln x + (N+1)\ln z + \sum_{\nu = 0}^{\infty}
    g_{\nu}\ln\!\left(1 - z x^{\nu}\right) \; .
\label{MFF}
\end{equation}
The saddle-point $(x_0,z_0)$ for the double contour integral~(\ref{DCI})
now is determined by the simultaneous solution of the two equations     
\begin{equation}
    \left. \frac{\partial \oF(x,z)}{\partial x} \right|_{x=x_0, z=z_0} = 0
    \qquad , \qquad
    \left. \frac{\partial \oF(x,z)}{\partial z} \right|_{x=x_0, z=z_0} = 0
    \; ,
\end{equation}
reading explicitly 
\begin{eqnarray}
    m + 1 & = & \sum_{\nu=0}^{\infty} g_{\nu}\frac{\nu}{z_0^{-1}x_0^{-\nu}-1}
\nonumber \\
    N + 1 & = & \sum_{\nu=0}^{\infty} g_{\nu}\frac{1}{z_0^{-1}x_0^{-\nu}-1}
    \; .
\label{MSE}
\end{eqnarray}
As long as the gas is not condensed, we may safely use the standard
approximation scheme~\cite{GajdaRzazewski97}. That is, we may expand the
function~(\ref{MFF}) quadratically around the saddle-point,
\begin{equation}
    \oF(x,z) \approx \oF(x_0,z_0) 
    + \frac{1}{2}\oF^{(2,0)}(x-x_0)^2
    + \oF^{(1,1)}(x-x_0)(z-z_0) +\frac{1}{2}\oF^{(0,2)}(z-z_0)^2
\end{equation}
with
\[
    \oF^{(r,s)} \equiv \left.
    \frac{\partial^{\,r+s}\oF(x,z)}{\partial x^r \partial z^s}
    \right|_{x=x_0, z=z_0} \; ,
\]
substitute $x - x_0 = iw$ and $z - z_0 = iu$, and get
\begin{eqnarray}
    \Omega(m,N) & \sim & 
    \frac{1}{(2\pi)^2}\exp\!\left(-\oF(x_0,z_0)\right)
\nonumber \\ & & \quad \times 
    \int_{-\infty}^{+\infty} \! {\rm d}w
    \int_{-\infty}^{+\infty} \! {\rm d}u \,
    \exp\!\left(\frac{1}{2}\oF^{(2,0)} w^2 + \oF^{(1,1)} wu
    + \frac{1}{2}\oF^{(0,2)} u^2 \right)
\nonumber \\
    & = & \frac{1}{2\pi\sqrt{D}}\exp\!\left(-\oF(x_0,z_0)\right) \; ,
\end{eqnarray}
where $D$ is the functional determinant
\begin{equation}
    D = {\rm det} \! \left( \begin{array}{cc}
	\oF^{(2,0)} & \oF^{(1,1)}   \\
	\oF^{(1,1)} & \oF^{(0,2)} \end{array} \right)
    = \oF^{(2,0)} \oF^{(0,2)} - \left(\oF^{(1,1)}\right)^2 \; .	
\end{equation}
For temperatures below the onset of condensation, however, this procedure is
invalid, because then the second of the saddle-point equations~(\ref{MSE})
dictates that $z_0$ differs from the ground-state singularity $z = 1$ of the
grand canonical partition function~(\ref{DSE}) merely by a quantity of order
$O(1/N)$, exactly as in the canonical case. Hence, we have to proceed 
according to the insight accumulated there for computing $\Omega(m,N)$ in
the condensate regime, and have to exempt the ground-state factor from the
the quadratic expansion: Defining 
\begin{equation}
    F(x,z) = \oF(x,z) - \ln(1-z) \; ,
\end{equation}
we have to start from the representation
\begin{equation} 
    \Omega(m,N) = 
    \frac{1}{(2\pi i)^2} \oint \! {\rm d}x \oint \! {\rm d}z \,
    \frac{\exp\!\left(-F(x,z)\right)}{1-z} \; .	
\end{equation}
The expansion of the tempered function $F(x,z)$ around $(x_0,z_0)$ then
yields
\begin{eqnarray}
    F(x,z) & \approx & F(x_0,z_0) + \frac{1}{2}F^{(2,0)}(x-x_0)^2
\nonumber \\ & &
    + \left[ F^{(0,1)} + F^{(1,1)}(x-x_0) \right]\!(z-z_0)
    + \frac{1}{2}F^{(0,2)}(z-z_0)^2 \; ,
\label{QUA}
\end{eqnarray}	
where we have used
\begin{eqnarray}
    F^{(1,0)} = 0 \; .
\end{eqnarray}
It is now of key interest to compare the magnitude of the two terms in the
square brackets of Eq.~(\ref{QUA}), which constitute the coefficient of
$(z - z_0)$. On the one hand, we have 
\begin{equation}
    F^{(0,1)} =
    -\left. \frac{\partial}{\partial z} \ln(1-z) \right|_{z=z_0}
    = \frac{1}{1 - z_0} \; ;
\end{equation}
hence $F^{(0,1)} = O(N)$ in the condensate regime. On the other hand,
relevant contributions to the integral over~$x$ are collected from an
interval of order $O\!\left(1/\sqrt{-F^{(2,0)}}\right)$ around $x_0$. Thus, 
the relevant $F^{(1,1)}(x-x_0)$ are on the order of
$F^{(1,1)}/\sqrt{-F^{(2,0)}}$, with
\begin{eqnarray}
    F^{(1,1)} & = & -\sum_{\nu=1}^{\infty} g_{\nu}
    \frac{\nu x_0^{\nu-1}}{\left(1 - z_0 x_0^{\nu}\right)^2} \; ,
\nonumber \\
    F^{(2,0)} & = & -\sum_{\nu=1}^{\infty} g_{\nu}
    \frac{\nu^2 z_0 x_0^{\nu-2}}{\left(1 - z_0 x_0^{\nu}\right)^2} \; .
\end{eqnarray}
If we focus again on systems of the type~(\ref{GRO}) --- i.e., if
both the quantum $\hbar\omega$ and the weights $g_{\nu}$ have been
adjusted accordingly ---, we may repeat the reasoning that has led to
the canonical estimate~(\ref{SUM}), and conclude that both $F^{(1,1)}$
and $F^{(2,0)}$ are of the order $O(N^{2s/d})$. Therefore,
$F^{(1,1)}(x-x_0) = O(N^{s/d})$ for relevant~$x$. To be honest, this can
be taken as a rather crude guideline only, in the same sense as the
quantities~$r_n$ displayed in Fig.~\ref{F_1} have not yet approached the
expected value $1/3$ for $n=2$. In fact, numerical calculations for the
three-dimensional isotropic harmonic oscillator potential reveal that for
reasonably large~$N$ the expression $F^{(1,1)}/\sqrt{-F^{(2,0)}}$ is about
proportional to $\langle N_{\rm ex} \rangle_{\rm mc}^{1/2}$, the square root
of the total number of excited particles in a microcanonical setting, instead
of being proportional to $\langle N_{\rm ex} \rangle_{\rm mc}^{1/3}$.
Nonetheless, the above estimate indicates that for $d/s > 1$ and large~$N$
we may neglect $F^{(1,1)}(x-x_0)$ against $F^{(0,1)}$. This implies a drastic
simplification of the analysis, because then the remaining saddle-point
integral factorizes: Leading both contours parallel to the respective
imaginary axis over the saddle, we are left with    
\begin{eqnarray}
    \Omega(m,N) & \sim & \frac{\exp\!\left(-F(x_0,z_0)\right)}{2\pi i}
    \int_{x_0-i\infty}^{x_0+i\infty} \! {\rm d}x \,
    \exp\!\left(-\frac{1}{2}F^{(2,0)}(x-x_0)^2\right)
\nonumber \\ & & \times
    \frac{1}{2\pi i} \int_{z_0-i\infty}^{z_0+i\infty} \! {\rm d}z \,
    \frac{
    \exp\!\left(-F^{(0,1)}(z-z_0)-\frac{1}{2}F^{(0,2)}(z-z_0)^2\right)
    }{1-z} \; .
\end{eqnarray}
The first of these integrals is standard, the second is precisely of
the type worked out in Appendices~\ref{AA} and~\ref{AB}. Thus, without any
further labor we obtain
\begin{equation}
    \Omega(m,N) \sim \frac{\exp\!\left(-F(x_0,z_0)-1\right)}
                          {\sqrt{-2\pi F^{(2,0)}}} \; ,
\label{MPF}
\end{equation}
the saddle-point approximation to the microcanonical partition functions
in the condensate regime.

\subsection{Microcanonical occupation numbers and their fluctuations}

The computation of microcanonical occupation numbers $\namc$ in the
condensate regime is a matter of routine by now, so we merely need to
sketch the main steps. Denoting the number of microstates of an isolated
$N$-particle system with total excitation energy $m\hbar\omega$ and with
exactly $n_\alpha$ particles occupying a given single-particle state with
energy $\alpha\hbar\omega$ as $\Gamma_\alpha(n_\alpha;m,N)$, we have
\begin{equation}
    \sum_{n_\alpha=0}^{N} \Gamma_\alpha(n_\alpha;m,N) = \Omega(m,N)
\end{equation}
and
\begin{equation}
    \namc = \frac{\sum_{n_\alpha=0}^N n_\alpha \Gamma_\alpha(n_\alpha;m,N)}
                 {\Omega(m,N)} \; .
\label{RAT}
\end{equation}
Introducing the symbol $\overline\partial/\partial(x^\alpha)$, where the
overbar is meant to indicate that the partial derivative acts on only
{\em one\/} of the $g_\alpha$-fold degenerate states with energy
$\alpha\hbar\omega$, the first microcanonical moments are generated from
the grand canonical partition function~(\ref{DSE}) by means of the identity
\begin{equation}
    x^\alpha \frac{\overline\partial}{\partial(x^\alpha)} \, \Xi(x,z)
    = \sum_{N=0}^{\infty} z^N \sum_{m=0}^{\infty} x^m
    \left(\sum_{n_\alpha=0}^N n_\alpha \Gamma_\alpha(n_\alpha;m,N)\right)
    \; .
\end{equation}
Hence,
\begin{eqnarray}
    \sum_{n_\alpha=0}^N n_\alpha \Gamma_\alpha(n_\alpha;m,N) & = &
    \frac{1}{(2\pi i)^2} \oint \! {\rm d}x \oint \! {\rm d}z \,
    \frac{x^\alpha}{x^{m+1}z^{N+1}}
    \frac{\overline\partial}{\partial(x^\alpha)} \, \Xi(x,z)
\nonumber \\ & \equiv &
    \frac{1}{(2\pi i)^2} \oint \! {\rm d}x \oint \! {\rm d}z \,	
    \exp\!\left(-\oG(x,z\right) \; ,
\label{MON}	
\end{eqnarray}
where
\begin{equation}
    \oG(x,z) = (m+1-\alpha)\ln x + N \ln z
    + \sum_{\nu = 0}^{\infty} g_\nu \ln\!\left(1 - zx^\nu\right)
    + \ln\!\left(1 - zx^\alpha\right) \; .
\end{equation}
The evaluation of the integral~(\ref{MON}) first requires the knowledge
of its saddle-point $(x_1,z_1)$, obtained by simultaneously solving the
two equations  
\begin{eqnarray}
    m + 1 - \alpha & = &
    \sum_{\nu=0}^{\infty} g_{\nu}\frac{\nu}{z_1^{-1}x_1^{-\nu}-1}
    + \frac{\alpha}{z_1^{-1}x_1^{-\alpha}-1} \; ,
\nonumber \\
    N & = & \sum_{\nu=0}^{\infty} g_{\nu}\frac{1}{z_1^{-1}x_1^{-\nu}-1}
    + \frac{1}{z_1^{-1}x_1^{-\alpha}-1} \; .
\end{eqnarray}
If then $\alpha \neq 0$, we define the tempered function
\begin{equation}
    G(x,z) = \oG(x,z) - \ln(1-z) \; ,
\end{equation}
yielding
\begin{eqnarray}
    \sum_{n_\alpha=0}^N n_\alpha \Gamma_\alpha(n_\alpha;m,N) & = &     
    \frac{1}{(2\pi i)^2} \oint \! {\rm d}x \oint \! {\rm d}z \,
    \frac{\exp\!\left(-G(x,z)\right)}{1-z}
\nonumber \\ & \sim &
    \frac{\exp\!\left(-G(x_1,z_1)-1\right)}{\sqrt{-2\pi G^{(2,0)}}}
\label{MFA}
\end{eqnarray}
in direct analogy to Eq.~(\ref{MPF}), with			      
\begin{equation}
    G^{(2,0)} = -\sum_{\nu=1}^\infty g_{\nu}
    \frac{\nu^2 z_1 x_1^{\nu-2}}{(1- z_1 x_1^{\nu})^2}
    - \frac{\alpha^2 z_1 x_1^{\alpha-2}}{(1 - z_1 x_1^{\alpha})^2} \; .
\end{equation}
If, however, $\alpha = 0$, we have to account for ground-state doubling.
In this case we define
\begin{equation}
    G(x,z) = \oG(x,z) - 2\ln(1-z)
\end{equation}
and invoke Eq.~(\ref{LTA}) with $\sigma=2$, resulting in
\begin{eqnarray}
    \sum_{n_0=0}^N n_0 \, \Gamma_0(n_0;m,N) & = &
    \frac{1}{(2\pi i)^2} \oint \! {\rm d}x \oint \! {\rm d}z \,
    \frac{\exp\!\left(-G(x,z)\right)}{(1-z)^2}
\nonumber \\ & \sim &
    \frac{2}{1-z_1}
    \frac{\exp\!\left(-G(x_1,z_1)-2\right)}{\sqrt{-2\pi G^{(2,0)}}} \; .
\label{MFM}
\end{eqnarray}
To give at least one application of these formulas, Fig.~\ref{F_10} shows
the microcanonical ground-state occupation number $\nzmc$ as a function
of the microcanonical temperature for $N = 1000$ ideal Bose particles in an
isotropic harmonic oscillator potential, as computed from Eqs.~(\ref{MPF})
and~(\ref{MFM}) according to Eq.~(\ref{RAT}). The microcanonical temperature
does not differ significantly from the canonical one~\cite{GajdaRzazewski97};
the inset quantifies the ratio of canonical to grand canonical, and of 
microcanonical to grand canonical occupation numbers. As expected, the
relative differences between the ground-state occupation numbers in the
three ensembles are on the order of $1/N$.  

For calculating the corresponding microcanonical mean-square fluctuations
$\famc$ with\-in the saddle-point approximation, we exploit the identity
\begin{eqnarray}
    \left( x^\alpha \frac{\overline\partial}{\partial(x^\alpha)} \right)^2
    \Xi(x,z) & = &
    x^\alpha \frac{\overline\partial}{\partial(x^\alpha)} \, \Xi(x,z)
    + x^{2\alpha} \frac{\overline\partial^2}{\partial(x^\alpha)^2} \, \Xi(x,z)
\nonumber \\ & = & 
    \sum_{N=0}^{\infty} z^N \sum_{m=0}^{\infty} x^m
    \left(\sum_{n_\alpha=0}^N n_\alpha^2 \Gamma_\alpha(n_\alpha;m,N)\right)
\end{eqnarray}
which immediately leads to the analogue of Eq.~(\ref{COF}), namely
\begin{equation}
    \famc = \namc - \namc^2 + \frac{2}{\Omega(m,N)} \,
    \frac{1}{(2\pi i)^2} \oint \! {\rm d}x \oint \! {\rm d}z \,
    \exp\!\left(-\oH(x,z)\right) \; .
\label{ANA}
\end{equation} 
The newly appearing integral is defined by
\begin{eqnarray} & &
    \frac{1}{(2\pi i)^2} \oint \! {\rm d}x \oint \! {\rm d}z \,
    \exp\!\left(-\oH(x,z\right)
\nonumber \\ & \equiv &	
    \frac{1}{(2\pi i)^2} \oint \! {\rm d}x \oint \! {\rm d}z \,
    \frac{\frac{1}{2}x^{2\alpha}}{x^{m+1}z^{N+1}}
    \frac{\overline\partial^2}{\partial(x^\alpha)^2} \, \Xi(x,z) \; ,
\end{eqnarray}
giving
\begin{equation}
    \oH(x,z) = (m+1-2\alpha)\ln x + (N-1) \ln z
    + \sum_{\nu = 0}^{\infty} g_\nu \ln\!\left(1 - zx^\nu\right)
    + 2\ln\!\left(1 - zx^\alpha\right) \; .
\label{MFH}
\end{equation}
Hence, its saddle-point $(x_2,z_2)$ is found by simultaneously solving
the two equations 
\begin{eqnarray}
    m + 1 - 2\alpha & = &
    \sum_{\nu=0}^{\infty} g_{\nu}\frac{\nu}{z_2^{-1}x_2^{-\nu}-1}
    + \frac{2\alpha}{z_2^{-1}x_2^{-\alpha}-1} \; ,
\nonumber \\
    N - 1 & = & \sum_{\nu=0}^{\infty}
    g_{\nu}\frac{1}{z_2^{-1}x_2^{-\nu}-1}
    + \frac{2}{z_2^{-1}x_2^{-\alpha}-1} \; .
\end{eqnarray}
The usual distinction follows: If $\alpha \neq 0$, extracting the
ground-state contribution from the function~(\ref{MFH}) means
introducing
\begin{equation}
    H(x,z) = \oH(x,z) - \ln(1-z) \; ,
\end{equation}
resulting in
\begin{equation}
    \frac{1}{(2\pi i)^2} \oint \! {\rm d}x \oint \! {\rm d}z \,
    \frac{\exp\!\left(-H(x,z)\right)}{1-z} \sim
    \frac{\exp\!\left(-H(x_2,z_2)-1\right)}{\sqrt{-2\pi H^{(2,0)}}}
\label{MSA}
\end{equation}
with
\begin{equation}
    H^{(2,0)} = -\sum_{\nu=1}^\infty g_{\nu}
    \frac{\nu^2 z_2 x_2^{\nu-2}}{(1- z_2 x_2^{\nu})^2}
    - 2\frac{\alpha^2 z_2 x_2^{\alpha-2}}{(1 - z_2 x_2^{\alpha})^2} \; .
\end{equation}
If $\alpha = 0$, we define instead
\begin{equation}
    H(x,z) = \oH(x,z) - 3\ln(1-z) \; ,
\end{equation}
and invoke Eq.~(\ref{LTA}) once more, now for $\sigma = 3$, to arrive at
\begin{equation}
    \frac{1}{(2\pi i)^2} \oint \! {\rm d}x \oint \! {\rm d}z \,
    \frac{\exp\!\left(-H(x,z)\right)}{(1-z)^3} \sim
    \frac{1}{2} \frac{9}{(1-z_2)^2}
    \frac{\exp\!\left(-H(x_2,z_2)-3\right)}{\sqrt{-2\pi H^{(2,0)}}} \; .
\label{MSM}
\end{equation}
Collecting the results~(\ref{MPF}), (\ref{MFA}), and (\ref{MSA}) for
$\alpha \neq 0$, or (\ref{MPF}), (\ref{MFM}), and (\ref{MSM}) for
$\alpha = 0$, Eq.~(\ref{ANA}) then allows one to determine the fluctuations.
An example for such a calculation is depicted in Fig.~\ref{F_11}: The heavy
solid line is the root-mean-square fluctuation $\mzmc \equiv \fzmc^{1/2}$ as
obtained from the above saddle-point scheme for a gas of $10^6$ ideal Bosons
in the usual isotropic oscillator potential. For comparison, when evaluating
for the same system the formula~(\ref{IDF}) for the difference $\fzcn - \fzmc$
up to terms of the order $k_BT/(\hbar\omega)$, one finds
\begin{eqnarray}
    \fzcn & - & \fzmc
\nonumber \\
    & = & \left(\frac{k_BT}{\hbar\omega}\right)^3
    \frac{3}{4} \frac{\zeta^2(3)}{\zeta(4)}
\nonumber \\
    & + & \left(\frac{k_BT}{\hbar\omega}\right)^2
    \left[\frac{3}{2} \frac{\zeta(2)\zeta(3)}{\zeta(4)} -
          \frac{9}{16} \frac{\zeta^3(3)}{\zeta^2(4)}\right]
\nonumber \\
    & + & \;\; \frac{k_BT}{\hbar\omega} \frac{\zeta(3)}{2\,\zeta(4)}
    \left[\ln\!\left(\frac{k_BT}{\hbar\omega}\right) + \gamma + \frac{5}{24}
          + \frac{27}{32} \frac{\zeta^3(3)}{\zeta^2(4)}
	  - \frac{5}{2} \frac{\zeta(2)\zeta(3)}{\zeta(4)}
	  + \frac{3}{2} \frac{\zeta^2(2)}{\zeta(3)}\right] \; .
\label{DOF}
\end{eqnarray}  
This, together with Eq.~(\ref{OAP}), yields a closed expression for the
microcanonical condensate fluctuations of an ideal Bose gas in a
three-dimensional isotropic harmonic trap in the oscillator approximation,
that is, under the assumption of an infinite reservoir of condensate
particles:  
\begin{eqnarray}
    \fzmc \approx 0.64366 \left(\frac{k_BT}{\hbar\omega}\right)^3
    & + & 
    \left[1.5 \ln\!\left(\frac{k_BT}{\hbar\omega}\right) + 1.85443 \right]
    \left(\frac{k_BT}{\hbar\omega}\right)^2
\nonumber \\
    & - & 
    \left[0.55531 \ln\!\left(\frac{k_BT}{\hbar\omega}\right) + 0.96969 \right]
    \frac{k_BT}{\hbar\omega} \; .
\label{MFL}
\end{eqnarray}
The dashed line in Fig.~\ref{F_11} corresponds to the leading term of this
approximation~\cite{NavezEtAl97}, whereas the thin line also takes the
next-to-leading order into account. Even for $10^6$ particles, finite-size
effects are still visible in the condensate fluctuation; the leading-order
term of Eq.~(\ref{MFL}) alone yields only modest agreement with the
saddle-point result. After accounting for the dominant corrections, the
agreement becomes close to perfect: Even on the scale of the inset, the
saddle-point result is indistinguishable from the oscillator
approximation~(\ref{MFL}).

\section{Discussion}
\label{S4}

The necessity to abandon the usual saddle-point scheme when exploring
canonical or micro\-canonical statistics of condensed Bose gases with
$N$~particles is brought about by a characteristic dilemma. On the one
hand, the approach of the saddle-point to the ground-state singularity at
$z = e^{\beta\varepsilon_0}$ of the grand canonical partition function
within order $O(1/N)$ may be taken as the very hallmark of Bose--Einstein
condensation; on the other, the customary Gaussian approximation requires
that intervals of order $O(1/N)$ around the saddle-point stay clear of
singularities. The solution to this problem almost suggests itself: If one
exempts the ground-state factor of the grand canonical partition function
from the Gaussian expansion and treats that factor exactly, but proceeds as
usual otherwise, then the singular point that now decides the fate of the
approximation is the one produced by the first excited state at
$z = e^{\beta\varepsilon_1}$. Since the saddle-point remains pinned below
$e^{\beta\varepsilon_0}$, it remains separated from the decisive singularity
at $z = e^{\beta\varepsilon_1}$ by an $N$-independent gap. This gap is 
wide enough to get the approximation going if the particle number is
sufficiently large, because the required interval of regularity shrinks
with increasing~$N$. The representation~(\ref{CPF}) of the canonical
partition functions can be viewed as the prototype integral expressing this
strategy; the other canonical and microcanonical quantities computed in this
work constitute nothing but variations of the same mathematical theme. 

The success of this amended saddle-point method hinges on the fact that
the emerging integrals with singular integrands can be done exactly; as
explained in Appendix~\ref{AA}, they lead directly to parabolic cylinder
functions. Thus, we have accomplished the two goals set in the Introduction:
The results~(\ref{CSA}) are easy to use, and provide bona fide approximations
to partition functions, occupation numbers, and their fluctuations which are
valid at all temperatures --- not only in the high-temperature limit or in
the condensate regime, but also in the critical temperature range that
witnesses the onset of condensation. In particular, the sharpness of the
drop of the complementary error function contained in the canonical partition
function~(\ref{CEF}) allows one to precisely assess the sharpness of this
onset in a Bose gas with a finite, fixed number of particles. 

The observation that the interval of regularity claimed by the Gaussian
expansion around the saddle-point is of the same order $O(1/N)$ as the
distance of the saddle-point from the ground-state singularity, meaning that
the original conflict is not too large, reflects itself in the noteworthy
fact that the error of the conventional scheme in the condensate regime is
merely a temperature- and system-independent multiplicative constant. The
discovery, made in Appendix~\ref{AB}, that this constant approaches unity
when the multiplicity of the ground-state pole is increased fits in nicely:
Increasing that multiplicity amounts to considering a Bose gas with a
multiple-degenerate ground state and thus drives the saddle-point away from
the ground-state singularity --- the occupation number of each individual
of the degenerate states is lowered ---, thereby lessening the error of
the naive approach.

It is also of interest to compare the workload implied by the proper
saddle-point method to that required by other techniques aiming at
canonical or microcanonical statistics. Exact recursion relations like
Eq.~(\ref{FRF}) or their microcanonical analogues are invaluable for
treating relatively small samples with not substantially more than about
1000 particles~\cite{WeissWilkens97,BalazsBergeman98}, but the computation
of, e.g., microcanonical condensate fluctuations by such means for a gas
with $10^6$~Bosons, as presented in Fig.~\ref{F_11}, is entirely out of the
question. The integral representations~(\ref{ICN}) -- (\ref{IDF}), on the
other hand, immediately yield analytical expressions for condensate
occupation numbers and fluctuations, {\em provided\/} the pole structure of
the spectral Zeta function~(\ref{SZF}) is known, but they do not allow one
to monitor the system at the onset of condensation. In contrast, saddle-point
techniques always require one numerical step --- finding the saddle-point as
the root of the respective saddle-point equation ---, but once this has been
done, the formalism yields all quantities of interest, by means of the
ever-same formulas, without further hardship. Therefore, we may conclude that
despite all reservations~\cite{FujiwaraEtAl70,ZiffEtAl77} piled up in more
than half a century since Schubert's incisive comments~\cite{Schubert46},
it really is the saddle-point method which, if executed properly, provides
the most powerful approach to the statistical mechanics of isolated,
condensed ideal Bose gases.    

Having an instrument that reliable and flexible at one's disposal is certainly
not merely of mathematical value, but may also have some bearing on experiments
with Bose--Einstein condensates of dilute atomic vapors which are now
becoming routine. These experiments are mainly done in isolated harmonic
traps, in the regime of vapor densities where the atomic interactions,
quantified by the $s$-wave scattering length~$a$, can be considered as weak:
Denoting the atom mass as~$m$ and the characteristic trap frequency as
$\omega$, and defining the oscillator length $L = \sqrt{\hbar/(m\omega)}$
which quantifies the extension of the trap's ground state, one has
$N(a/L)^3 \ll 1$ under typical conditions. For example, $a = 5.4$~nm
for $^{87}$Rb~\cite{VogelsEtAl97}, giving $N(a/L)^3 = 8 \cdot 10^{-3}$
for a sample of $N = 10^6$ condensate atoms in a shallow trap with
$\omega = 100$~s$^{-1}$, while $Na/L = 2 \cdot 10^{3}$. These two relations
place the system in the Bogoliubov regime, traditionally associated with the
notion of a weakly interacting Bose gas. However, it is feasible to prepare
even more weakly interacting samples, either by tuning the scattering length
with the help of external magnetic fields~\cite{Inouye98}, or by working
with spin-polarized atomic hydrogen~\cite{FriedEtAl98}, which features the
unusually low triplet scattering length $a = 0.0648$~nm~\cite{JamiesonEtAl95}.
Thus, $a/L = 2.6 \cdot 10^{-6}$ in a harmonic trap with
$\omega = 100$~s$^{-1}$, so that even for $N \approx 400\,000$ hydrogen atoms
one finds $Na/L \approx 1$, indicating the crossover regime from the ideal to
the Bogoliubov gas. This crossover should manifest itself, in a non-trivial
manner, in the behavior of the condensate fluctuations~\cite{GiorginiEtAl98},
which also determine what one may aptly term ``the minimum linewidth of an
atom laser''~\cite{Graham98,Ketterle98}. It would therefore be of substantial
importance to study condensate fluctuations of {\em very weakly interacting
Bose gases\/}, that is, of systems intermediate between the ideal gas and the
Bogoliubov gas, and to probe whether the difference~(\ref{DOF}) between the
canonical and the microcanonical ensemble remains visible there;
in general, this difference should show a pronounced dependence on the trap
type~\cite{HKK98}. In this way, an old, apparently purely academic issue ---
the non-equivalence of statistical ensembles in the condensate regime ---
suddenly pops up at the forefront of topical research, in the theory of the
atom laser. Seen from the experimental angle, such an enterprise is on the
verge of becoming possible; on the theoretical side, the first requirement is
a tool for routinely computing ideal condensate fluctuations within the
different ensembles, for traps with various geometries. This tool is
available now.

\appendix

\section{Accurate saddle-point approximations for Bose-type integrals}
\label{AA}

In Section~\ref{S2} we have met contour integrals of the form
\begin{equation}
    I_\sigma \equiv \frac{1}{2\pi i} \oint \! {\rm d}z \,
    \frac{\exp\!\left(-f(z)-(\sigma-1)\beta\varepsilon_0\right)}
         {\left(1 - ze^{-\beta\varepsilon_0}\right)^\sigma} \; ,    
\label{ISI}
\end{equation}
with positive integer $\sigma$, and a saddle-point lying too close
to the singularity at $z = e^{\beta\varepsilon_0}$ for the standard
approximation~(\ref{WSA}) to be viable. In this appendix we derive the
proper saddle-point approximation to these integrals, following a
suggestion by Dingle~\cite{Dingle73}.

Writing, in accordance with our previous notation, the negative logarithm
of the full integrand as $\of(z)$, 
\begin{equation}
    \of(z) = f(z) + (\sigma-1)\beta\varepsilon_0 
    + \sigma\ln\!\left(1 - ze^{-\beta\varepsilon_0}\right) \; ,    
\end{equation}
the saddle-point $z_*$ is determined by the equation
\begin{equation}
    \left. \frac{{\rm d} \of(z)}{{\rm d} z} \right|_{z=z_*} = 0 \; .
\label{SPE}
\end{equation}
In the large-$N$-limit, this equation corresponds to the grand canonical
relation between particle number~$N$ and fugacity~$z_*$ for a system with
a $\sigma$-fold degenerate ground state.

Substituting $z = z_* + u$, so that the saddle-point is found at $u = 0$,
and writing the difference between the singular point and the saddle-point as
\begin{equation}
    u_0 \equiv e^{\beta\varepsilon_0} - z_* \; ,
\label{DOS}
\end{equation}
we have
\begin{equation}
    \of(z_*+u) = f(z_*+u) - \beta\varepsilon_0 + \sigma\ln(u_0 - u)
\label{FBU}
\end{equation}
and
\begin{equation}
    I_\sigma = \frac{e^{\beta\varepsilon_0}}{2\pi i} \oint \! {\rm d}u \,
    \frac{\exp\!\left(-f(z_*+u)\right)}{(u_0 - u)^\sigma} \; .
\end{equation}
Defining $f^{(n)} \equiv f^{(n)}(z_*)$, Eqs.~(\ref{SPE}) and (\ref{FBU})
immediately yield 
\begin{eqnarray}
    f^{(1)} & = & -\sigma \left. \frac{\rm d}{{\rm d}u} \ln(u_0-u)
    \right|_{u=0}
\nonumber \\
    & = & \frac{\sigma}{u_0} \; ;
\end{eqnarray}    
moreover, we require $f^{(2)} \ll 0$. Expanding $f(z_*+u)$ up to second order
around the saddle-point --- with a first derivative $f^{(1)}$ which does not,
as in the conventional approximation~(\ref{WSA}), vanish, but instead becomes
{\em large\/} when $u_0$ is small, as in the condensate regime ---, then
leading the path of integration over the saddle at $u=0$, we obtain the
approximation
\begin{eqnarray}
    2\pi i \, \Isp & = & (-1)^{\sigma}
    \exp\!\left(\beta\varepsilon_0 - f^{(0)}\right)
\nonumber \\ & & \times
    \int_{-i\infty}^{+i\infty} \! {\rm d}u \,
    \exp\!\left(-\frac{\sigma u}{u_0} -\frac{1}{2}f^{(2)}u^2\right)
    \left(u - u_0\right)^{-\sigma} \; ,
\end{eqnarray}
which, upon substituting
\begin{equation}
    u = i \frac{v}{\sqrt{-f^{(2)}}} + u_0 \; , 
\end{equation}
becomes
\begin{eqnarray}
    2\pi i \, \Isp & = & (-1)^{\sigma}
    \exp\!\left(\beta\varepsilon_0 - f^{(0)} - \sigma
         + \frac{1}{2}\eta^2\right)
    \left(\frac{\sqrt{-f^{(2)}}}{i}\right)^{\sigma-1}
\nonumber \\ & & \times
    \int_{-\infty+i\eta}^{+\infty+i\eta} \! {\rm d}v \,
    \exp\!\left(i\eb v -\frac{1}{2}v^2\right) v^{-\sigma} \; ,
\end{eqnarray}
with	
\begin{eqnarray}
    \eta   & \equiv & u_0\sqrt{-f^{(2)}}  
\label{XNB} \\
    \eb & \equiv & \eta - \frac{\sigma}{\eta} \; . 
\label{XWB}
\end{eqnarray}
The integral occurring here is closely related to Whittaker's parabolic
cylinder function $D_{-\sigma}(\eb)$, namely~\cite{ParabolicCF}
\begin{equation}
    \int_C \! {\rm d}v \,
    \exp\!\left(i\eb v -\frac{1}{2}v^2\right) v^{-\sigma}
    = \sqrt{2\pi} \, i^{-\sigma} e^{-\frac{1}{4}\eb^2} D_{-\sigma}(\eb) \; , 	
\end{equation}
where $C$ runs from $-\infty$ to $+\infty$, passing $v = 0$ from above.
Thus, the saddle-point approximation to the integrals~(\ref{ISI}) finally
takes the form 
\begin{eqnarray}
    \Isp & = & \frac{1}{\sqrt{2\pi}}\left(\sqrt{-f^{(2)}}\right)^{\sigma-1}
\nonumber \\ & & \times    
    \exp\!\left(\beta\varepsilon_0 - f^{(0)} - \sigma + \frac{1}{2}\eta^2
                - \frac{1}{4}\eb^2 \right) D_{-\sigma}(\eb) \; .
\label{CSA}
\end{eqnarray}

\section{High-- and low-temperature approximations for Bose-type integrals}
\label{AB}

The approximation~(\ref{CSA}) is valid regardless whether or not the
saddle-point $z_*$ lies close to $e^{\beta\varepsilon_0}$, that is, for all
temperatures. Therefore, it should adopt a more simple form in the condensate
regime, where $z_*^{-1}e^{\beta\varepsilon_0}-1$ is of the order $O(\sigma/N)$,
and should merge into the standard saddle-point formula in the
high-temperature limit, where $z_*$ approaches zero. 
 
In the case of high temperatures, the saddle-point moves away from the
ground-state singularity; the distance $u_0$ defined in Eq.~(\ref{DOS})
approaches unity, $u_0 \to 1$. Hence, both parameters~$\eta$ and
$\eb = \eta -\sigma/\eta$ introduced in Eqs.~(\ref{XNB}) and~(\ref{XWB})  
adopt large positive values. Then the asymptotic expansion of the parabolic
cylinder functions gives~\cite{WhittakerWatson62}  
\begin{equation}
    D_{-\sigma}(\eb) \sim \frac{\exp(-\eb^2/4)}{\eb^{\sigma}} \; ; 
\label{HTA}
\end{equation}
moreover, we have
\begin{equation}
    \frac{1}{2}\eta^2 - \frac{1}{2}\eb^2 \sim \sigma \; .
\label{HTS}
\end{equation}
Observing that, as a consequence of $u_0 \to 1$, 
\begin{eqnarray}
    \eb & \sim & \eta \sim \sqrt{-f^{(2)}}	\; ,
\\
    \of^{(0)} & \sim & f^{(0)} - \beta \varepsilon_0 \; ,
\\
    \of^{(2)} = f^{(2)} - \sigma u_0^{-2} & \sim & f^{(2)} \; , 
\label{HTF}
\end{eqnarray}
we find
\begin{eqnarray} 
    \Isp & \sim & \frac{1}{\sqrt{-2\pi f^{(2)}}}
    \exp(\beta\varepsilon_0 - f^{(0)})
\nonumber \\
    & \sim & \frac{\exp(-\of^{(0)})}{\sqrt{-2\pi\of^{(2)}}} \; . 
\label{USR}
\end{eqnarray}
This is just the expected result: For high temperatures, i.e., when $z_*$
stays sufficiently far away from $e^{\beta\varepsilon_0}$, we recover the
formula provided by the usual approximation scheme~(\ref{WSA}).

In the opposite limit, that is, in the condensate regime, we infer
$\sqrt{-f^{(2)}} = O(N^{\xi(2)/2})$ from Eq.~(\ref{ORD}), whereas $u_0$
is of order $O(\sigma/N)$. Since $\xi(2)/2 = \max\{1/2,s/d\}$ for systems
of the type~(\ref{GRO}), and we have required $s/d < 1$, we find that $u_0$
goes to zero faster than $\sqrt{-f^{(2)}}$ increases when~$N$ becomes large.
Hence, $\eta = u_0\sqrt{-f^{(2)}}$ approaches zero for large~$N$. This, in
turn, implies that the argument~$\eb$ of the parabolic cylinder functions now
is a large negative number, so that~\cite{WhittakerWatson62}
\begin{equation}
    D_{-\sigma}(\eb) \sim \frac{\sqrt{2\pi}}{(\sigma-1)!}(-\eb)^{\sigma-1}
    \exp(\eb^2/4) \; .
\end{equation}
Since
\begin{equation}
    -\eb \sim \frac{\sigma}{u_0\sqrt{-f^{(2)}}} \; ,
\end{equation}
the low-temperature limit of the approximation~(\ref{CSA}) becomes  
\begin{equation}
    \Isp \sim \frac{1}{(\sigma-1)!}\left(\frac{\sigma}{u_0}\right)^{\sigma-1}
    \exp\!\left(\beta\varepsilon_0 - f^{(0)} -\sigma\right) \; ,
\label{LTA}
\end{equation}
which has been used heavily in Secs.~\ref{S2} and~\ref{S3}.

It is of interest to recast this expression~(\ref{LTA}) into a form
which lends itself to a direct comparison with the now incorrect standard
saddle-point formula. To this end we exploit that, since $u_0 \to 0$, the
second derivative of the full function~$\of$ at the saddle-point will be
dominated by the singular ground-state contribution, giving 
\begin{equation}
    \of^{(2)} \sim -\frac{\sigma}{u_0^2} \; .
\end{equation}
With this additional approximation, and utilizing the identity
$\of^{(0)} = f^{(0)} -\beta\varepsilon_0 + \sigma \ln u_0$, 
Eq.~(\ref{LTA}) yields
\begin{eqnarray}
    \Isp & \sim & \frac{\sigma^{\sigma-1}}{(\sigma-1)!} \, u_0
    \exp\!\left(-\of^{(0)}-\sigma\right)
\nonumber \\
    & \sim & \frac{\sigma^{\sigma-1}e^{-\sigma}}{(\sigma-1)!} \,
    \sqrt{2\pi\sigma} \, \frac{\exp(-\of^{(0)})}{\sqrt{-2\pi\of^{(2)}}}
\nonumber \\
    & \equiv & R_{\sigma} \, \frac{\exp(-\of^{(0)})}{\sqrt{-2\pi\of^{(2)}}}
    \; .
\label{LTE}
\end{eqnarray}
This is a most intriguing observation: In the large-$N$, low-temperature
limit, i.e., in the condensate regime, the result of the properly executed
saddle-point approximation to the Bose-type integral~(\ref{ISI}) differs
from the standard saddle-point formula by a {\em temperature- and
system-independent\/} factor $R_\sigma$, namely    
\begin{equation}
    R_\sigma = \sqrt{2\pi\sigma} \, 
    \frac{\sigma^{\sigma-1} e^{-\sigma}}{(\sigma-1)!} \; ;
\end{equation}
some numerical values of $R_\sigma$ are listed in the following table.

\begin{center}
\begin{tabular}{|| c | c ||}	    \hline
    $\sigma$   &   $R_\sigma$	\\  \hline
        1      &    0.92214	\\    
        2      &    0.95950	\\
        3      &    0.97270	\\
        5      &    0.98349	\\
     ~~10~~    &  ~~0.99170~~	\\  \hline
\end{tabular}
\end{center}    

Recalling Stirling's formula for $(\sigma-1)!$, one immediately realizes
that these universal renormalization factors $R_\sigma$ approach unity
when the singularity index $\sigma$ is increased. This finding, which
might appear paradoxical at first, has a simple explanation: In a system
with a $\sigma$-fold degenerate ground state, each individual of these
states takes only $1/\sigma$-th of the particles that a non-degenerate
ground state would have to carry. Therefore,
$z_*^{-1}e^{\beta\varepsilon_0}-1$ is of the order $O(\sigma/N)$:
The larger $\sigma$, the farther away is the saddle-point $z_*$ from the
singularity at $z = e^{\beta\varepsilon_0}$, and the better is the standard
procedure.

On the other hand, in the high-temperature regime, where $u_0 \to 1$,
the left hand side of Eq.~(\ref{LTA}) can be written as
\begin{equation}
    \frac{\sigma^{\sigma-1} e^{-\sigma}}{(\sigma-1)!}
    \exp\!\left(-\of^{(0)}\right) \equiv
    M_\sigma \, \frac{\exp(-\of^{(0)})}{\sqrt{-2\pi\of^{(2)}}} \; .  
\end{equation}
Thus, when using the low-temperature approximation~(\ref{LTA}) in the
high-temperature regime, the result is incorrect by the factor
\begin{equation}
    M_\sigma = R_\sigma \sqrt{-\of^{(2)}/\sigma} \; .
\label{HTE}
\end{equation}

\begin{figure}
\caption[FIG.~1]{Circles: ratios~$r_n$, as defined by Eqs.~(\ref{SUM})
    and~(\ref{EXP}), for a gas of $N = 10^6$ ideal Bose particles with
    temperature $T = 0.5 \, T_0^{(3)}$ which is confined by an isotropic
    three-dimensional harmonic oscillator potential. According to
    Eq.~(\ref{SUM}), one expects $r_n \approx 1/3$ for largish~$n$.
    The diamonds indicate the corresponding data for a gas with same
    number of particles kept at $T = 0.5 \, T_0^{(1)}$ in a
    one-dimensional harmonic trap.}  
\label{F_1}
\end{figure}

\begin{figure}
\caption[FIG.~2]{Complementary error function ${\rm erfc}(\eb/\sqrt{2})$,
    with the temperature-dependent parameter $\eb$ defined by Eqs.~(\ref{CXN})
    and (\ref{CXW}), for a gas of $N = 10^3$ ideal Bose particles in a
    {\em three\/}-dimensional isotropic harmonic oscillator potential
    (full line; here the reference temperature~$T_0$ equals $T_0^{(3)}$ as
    given by Eq.~(\ref{TT3})), and in a {\em one\/}-dimensional harmonic
    potential (dashed line; with $T_0 = T_0^{(1)}$ as in Eq.~(\ref{TT1})).
    When ${\rm erfc}(\eb/\sqrt{2})$ approaches zero, the standard
    high-temperature result~(\ref{WSA}) holds; when it approaches two,
    Eq.~(\ref{COA}) provides the correct partition function.}
\label{F_2}
\end{figure}

\begin{figure}
\caption[FIG.~3]{Ratio of the standard saddle-point result~(\ref{WSA})
    (solid line approaching the value $1/R_1 \approx 1.08444$ at low
    temperatures), of the proper approximation~(\ref{CZN}) (solid line
    everywhere close to unity), and of its low-temperature
    descendant~(\ref{COA}) (dashed line), to the {\em exact\/} canonical
    partition function $Z_N(\beta)$, for a gas of $N = 1000$ ideal Bose
    particles in a three-dimensional iso\-tropic harmonic potential. The
    reference temperature~$T_0$ is given by Eq.~(\ref{TT3}). Note the
    impressive overall performance of the approximation~(\ref{COA}).}
\label{F_3}
\end{figure}

\begin{figure}
\caption[FIG.~4]{As Fig.~\ref{F_3}, now for a gas of $N = 1000$ ideal
    Bose particles in a one-dimensional harmonic potential. The reference
    temperature~$T_0$ is given by Eq.~(\ref{TT1}). Note that for $T/T_0 \to 0$
    the ratio of the standard approximation~(\ref{WSA}) and the exact data
    approaches the same value $1/R_1 \approx 1.08444$ as met in the
    three-dimensional case.}
\label{F_4}
\end{figure}

\begin{figure}
\caption[FIG.~5]{Canonical occupation number $\nocn$ as obtained from
    Eq.~(\ref{CSN}) (full line), and from the ratio of
    $\partial Z_N(\beta)/\partial(-\beta\varepsilon_1)$ to $Z_N(\beta)$,
    both computed from the standard saddle-point formula (dashed line),
    for $N = 1000$ ideal Bose particles in an isotropic three-dimensional
    harmonic potential. The inset shows the respective ratios of these
    approximate occupation numbers to the exact ones. The energy level
    $\varepsilon_1$ is three-fold degenerate; the data shown here
    correspond to an individual state. In this and all following figures,
    the reference temperature $T_0$ is given by Eq.~(\ref{TT3}).}  
\label{F_5}
\end{figure}

\begin{figure}
\caption[FIG.~6]{Canonical occupation number $\nzcn$ as obtained from
    Eq.~(\ref{CSZ}) (full line), and from the ratio of
    $\partial Z_N(\beta)/\partial(-\beta\varepsilon_0)$ to $Z_N(\beta)$,
    both computed from the standard saddle-point formula (dashed line),
    for the same system as considered in Fig.~\ref{F_5}. The inset shows
    the respective ratios of these approximate occupation numbers to the
    exact ones. The error of the standard approximation in the condensate
    regime is determined by the ratio $R_1/R_2 \approx 0.96106$; the error of
    the condensate approximation~(\ref{CSZ}) in the high-temperature regime
    is given by the factor $R_2/(R_1\sqrt{2}) \approx 0.73576$.}
\label{F_6}
\end{figure}

\begin{figure}
\caption[FIG.~7]{Canonical r.m.s.-fluctuation $\mocn$ as obtained from
    Eq.~(\ref{CFN}) (full line), and from the standard saddle-point
    approximations to the individual terms in Eq.~(\ref{COF}) (dashed line),
    for the same system as considered in Fig.~\ref{F_5}. The inset shows the
    respective ratios of these approximate fluctuations to the recursively
    computed exact one. As in Fig.~\ref{F_5}, the data refer to only one of
    the three states with energy $\varepsilon_1$.}
\label{F_7}
\end{figure}

\begin{figure}
\caption[FIG.~8]{Canonical r.m.s.-fluctuation $\mzcn$ as obtained from
    Eq.~(\ref{CFZ}) (full line), and from the standard saddle-point
    approximations to the individual terms in Eq.~(\ref{COF}) (dashed line),
    for the same system as considered in Fig.~\ref{F_5}. The inset shows the
    respective ratios of these approximate fluctuations to the recursively
    computed exact one. The error of the standard scheme for $T \to 0$
    is determined by Eq.~(\ref{ERR}), whereas the error of the condensate
    approximation for high temperatures is given by the factor
    $(R_2/(R_1\sqrt{2}))^{1/2} \approx 0.85776$.}
\label{F_8}
\end{figure}

\begin{figure}
\caption[FIG.~9]{Canonical r.m.s-fluctuation $\mzcn$ for the same system
    as considered in Fig.~\ref{F_5}. The exact, recursively computed
    fluctuation (long-dashed line) is compared to the data obtained
    from the saddle-point formula~(\ref{CFZ}) (heavy full line), to the   
    approximation provided by the leading, $T^3$-proportional term in
    Eq.~(\ref{OAP}) (short-dashed line), and to the prediction made by
    the oscillator approximation~(\ref{OAP}) with terms up to order~$T$
    (thin line). The inset emphasizes the outstanding accuracy of both the
    proper saddle-point method and the oscillator approximation.}
\label{F_9}
\end{figure}

\begin{figure}
\caption[FIG.~10]{Microcanonical ground-state occupation number $\nzmc$
    for a gas of $N = 1000$ ideal Bosons in a three-dimensional isotropic
    harmonic oscillator potential, as computed from the saddle-point
    approximations~(\ref{MPF}) and~(\ref{MFM}). The inset shows the ratio
    $\nzcn/\nzgc$ of the canonical to the grand canonical occupation numbers
    (short dashes; with exact, recursively computed canonical data), and the
    ratio $\nzmc/\nzgc$ of the microcanonical to the grand canonical values
    (long dashes).}    
\label{F_10}
\end{figure}

\begin{figure}
\caption[FIG.~11]{Microcanonical r.m.s.-condensate fluctuation $\mzmc$ for
    a gas of $N = 10^6$ ideal Bosons in a three-dimensional isotropic harmonic
    potential, as computed from the saddle-point approximation (full line).
    The dashed line corresponds to only the leading term of the oscillator
    approximation, cf.~Eq.~(\ref{MFL}); the thin line, visible only in the
    upper right corner, also takes the next-to-leading term into account.
    Even in the inset, the result of the saddle-point calculation is
    indistinguishable from this oscillator approximation.} 
\label{F_11}
\end{figure}


\begin{references}
\bibitem{ParabolicCF} This integral representation can be obtained
    from Eq.~(19.5.4) in Ref.~\cite{AbramowitzStegun72}, observing
    that the parabolic cylinder function $U(a,z)$ equals the
    Whittaker function $D_{-a-1/2}(z)$.
\end{references}
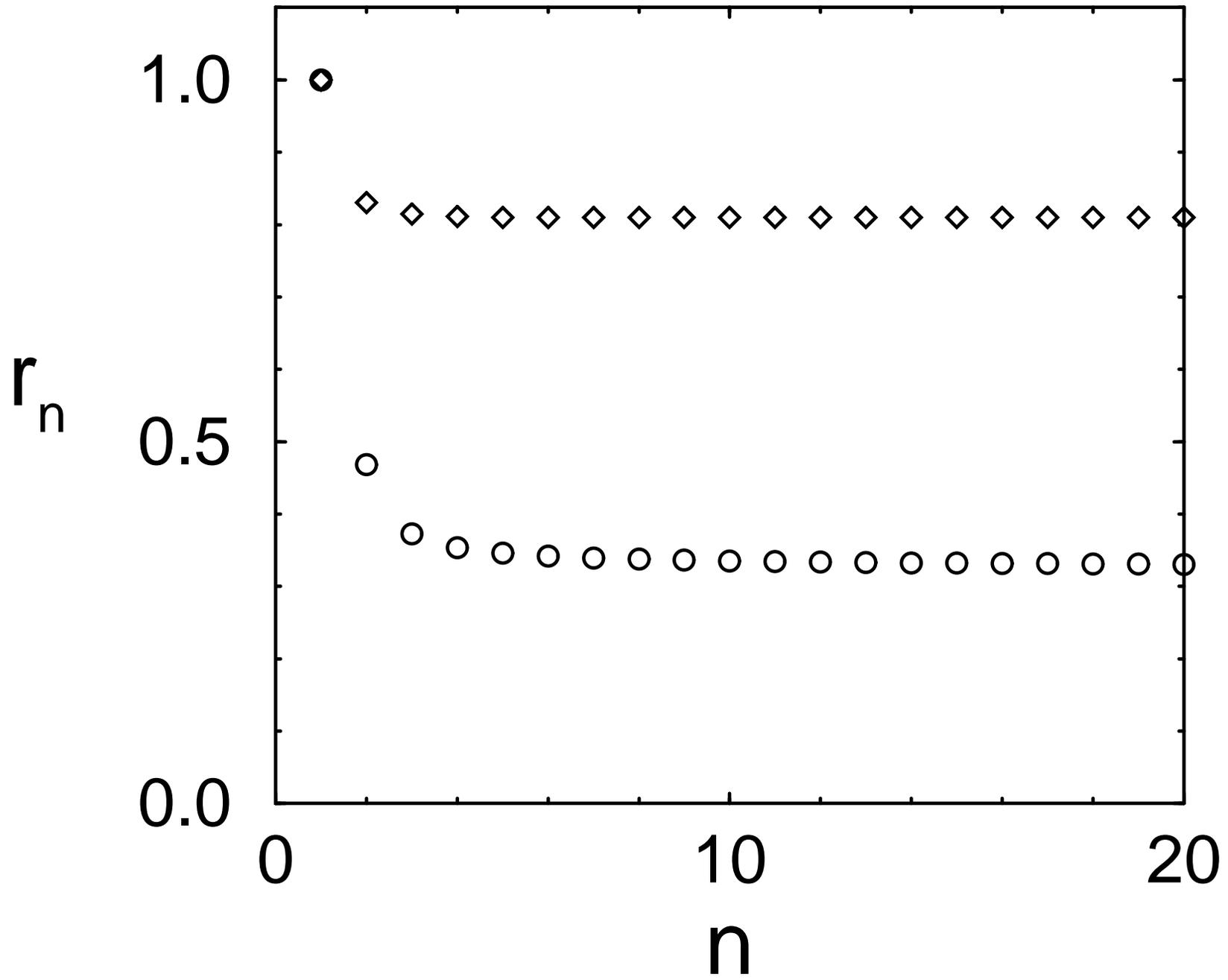
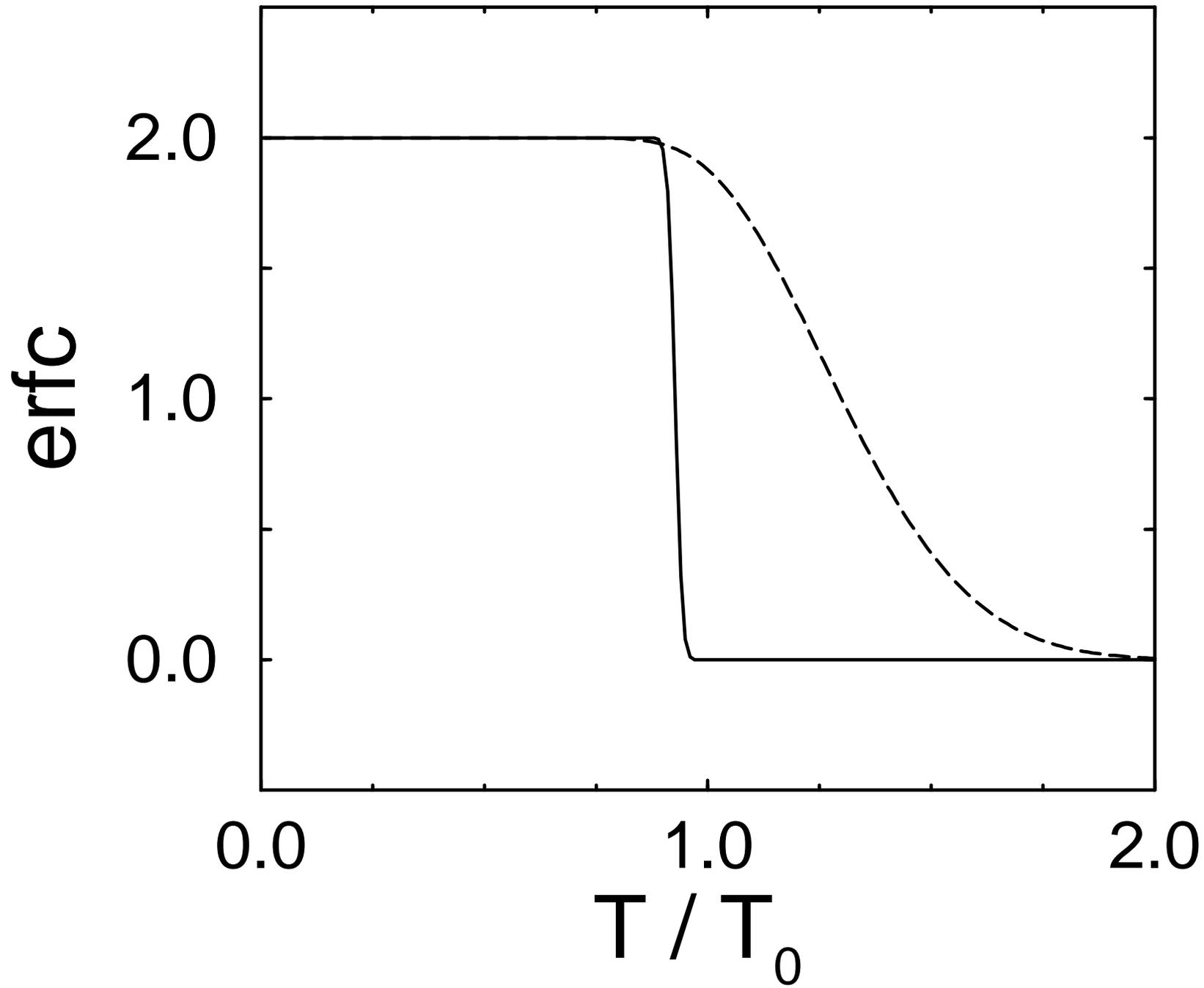
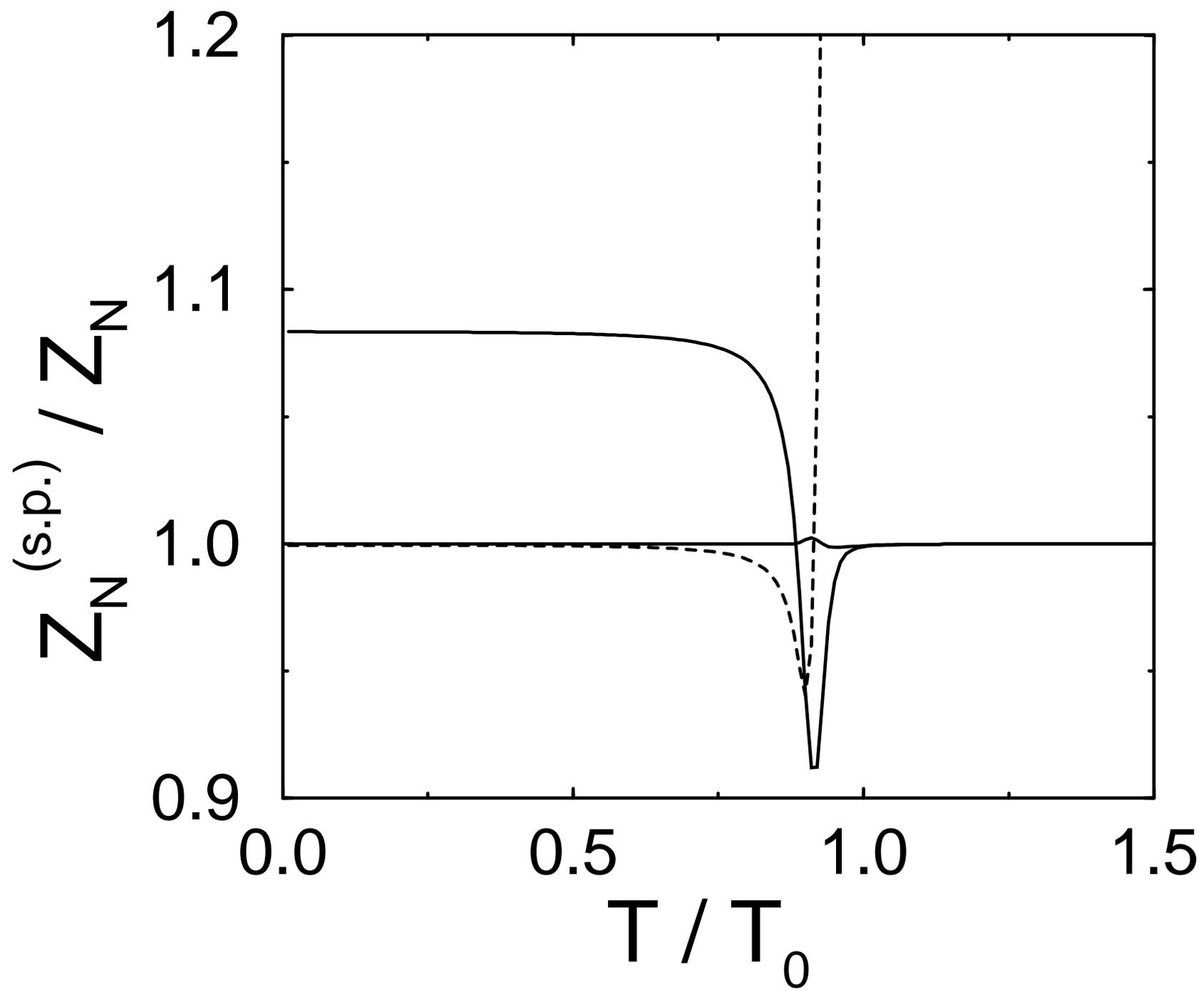
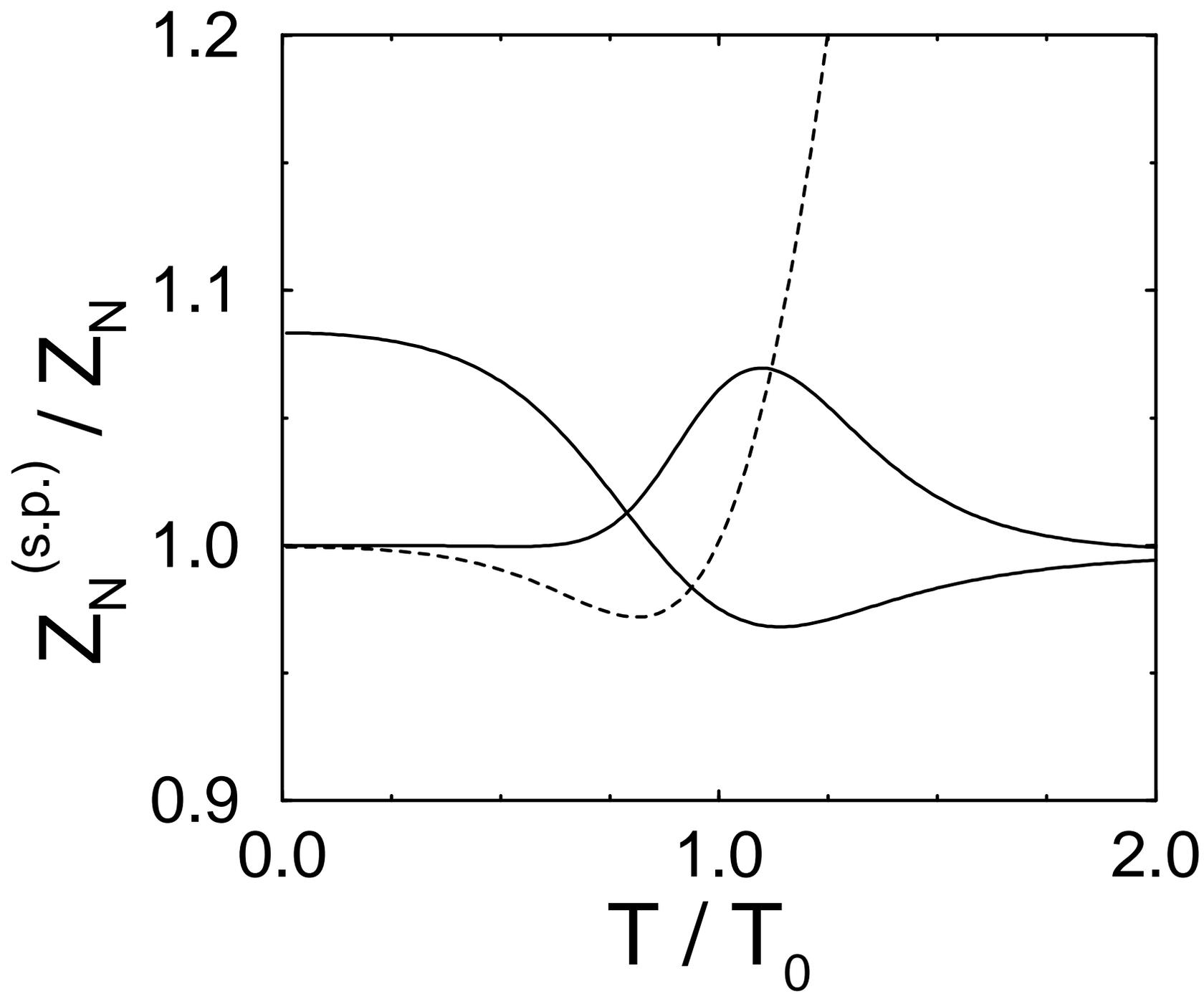
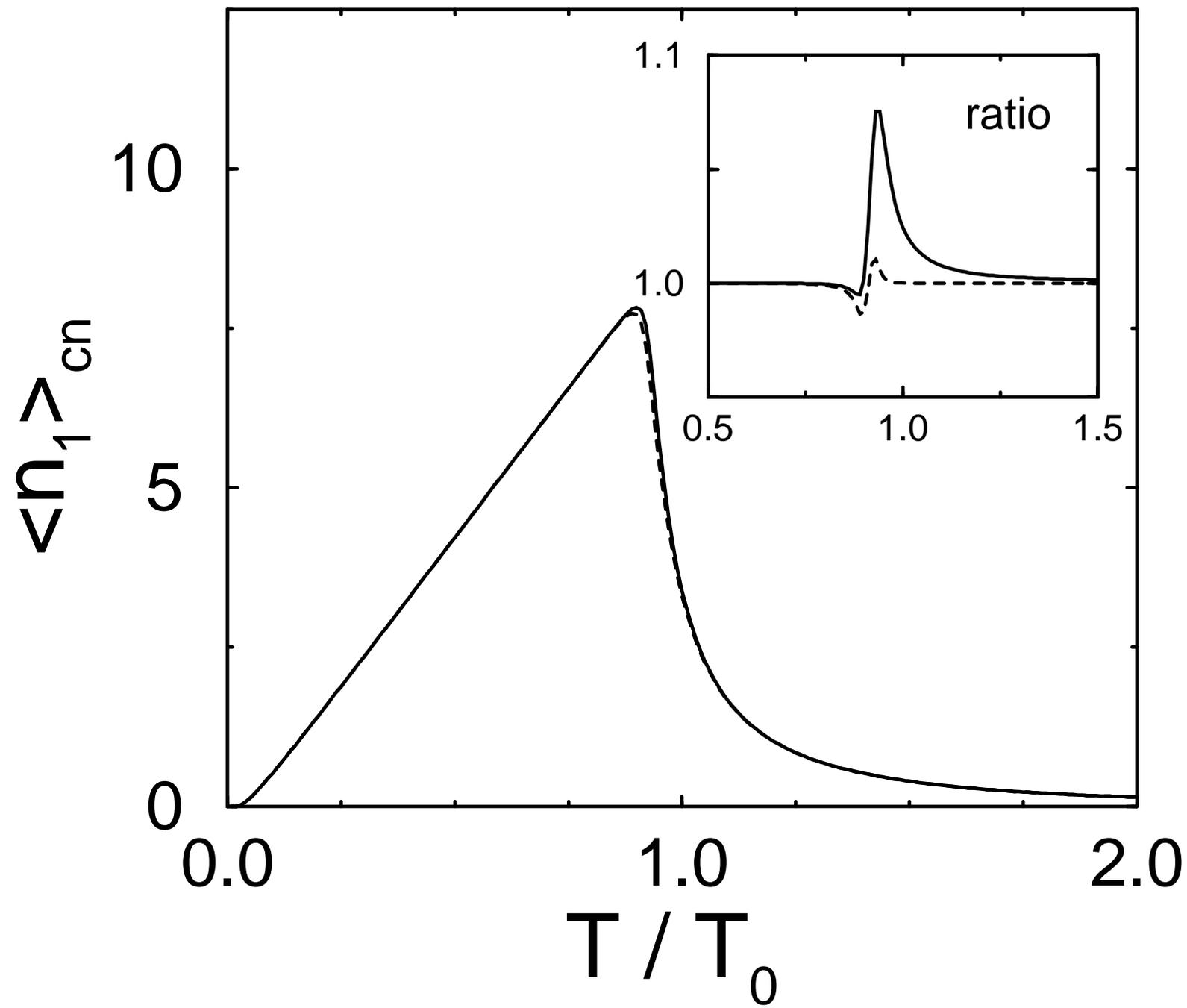
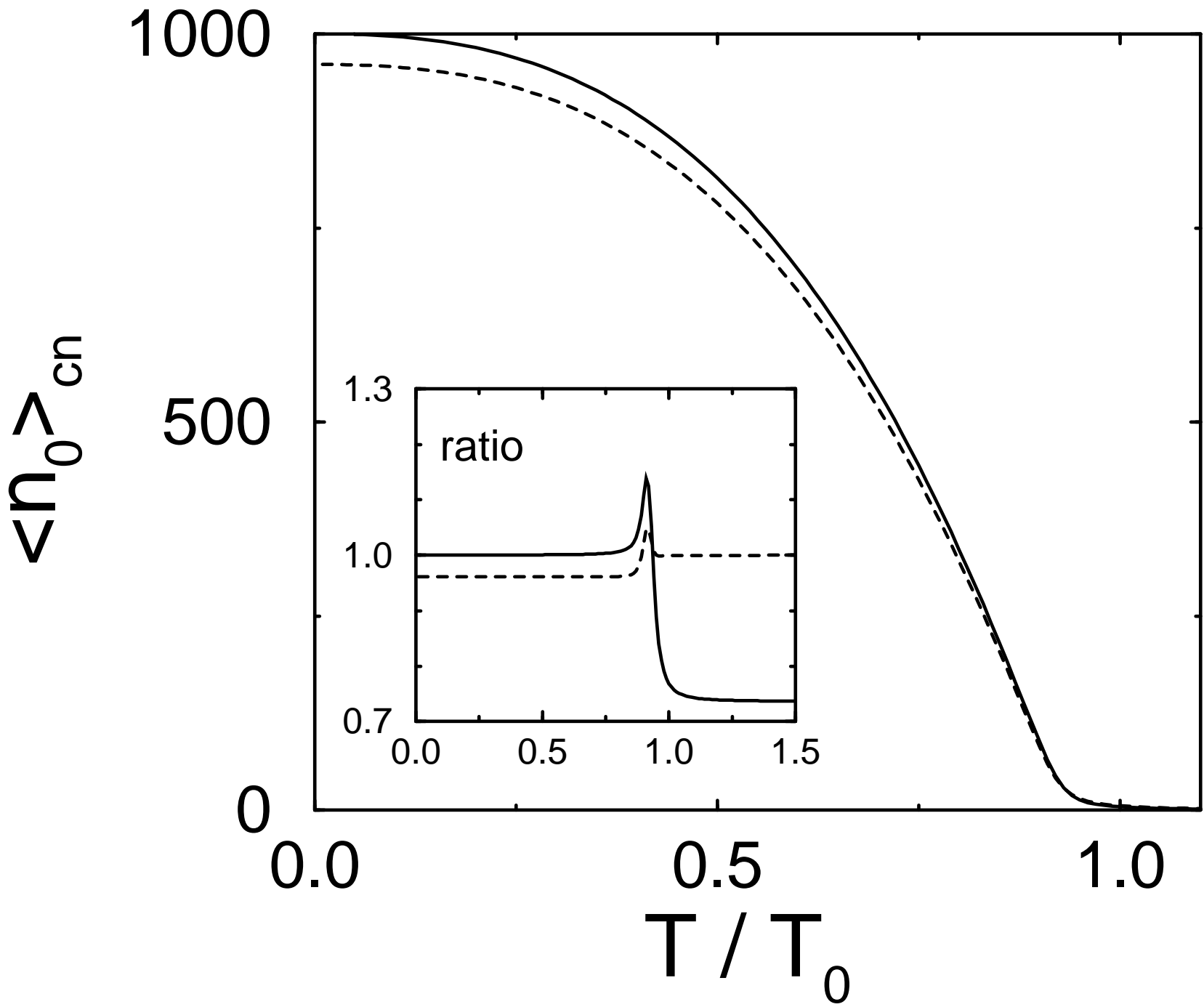
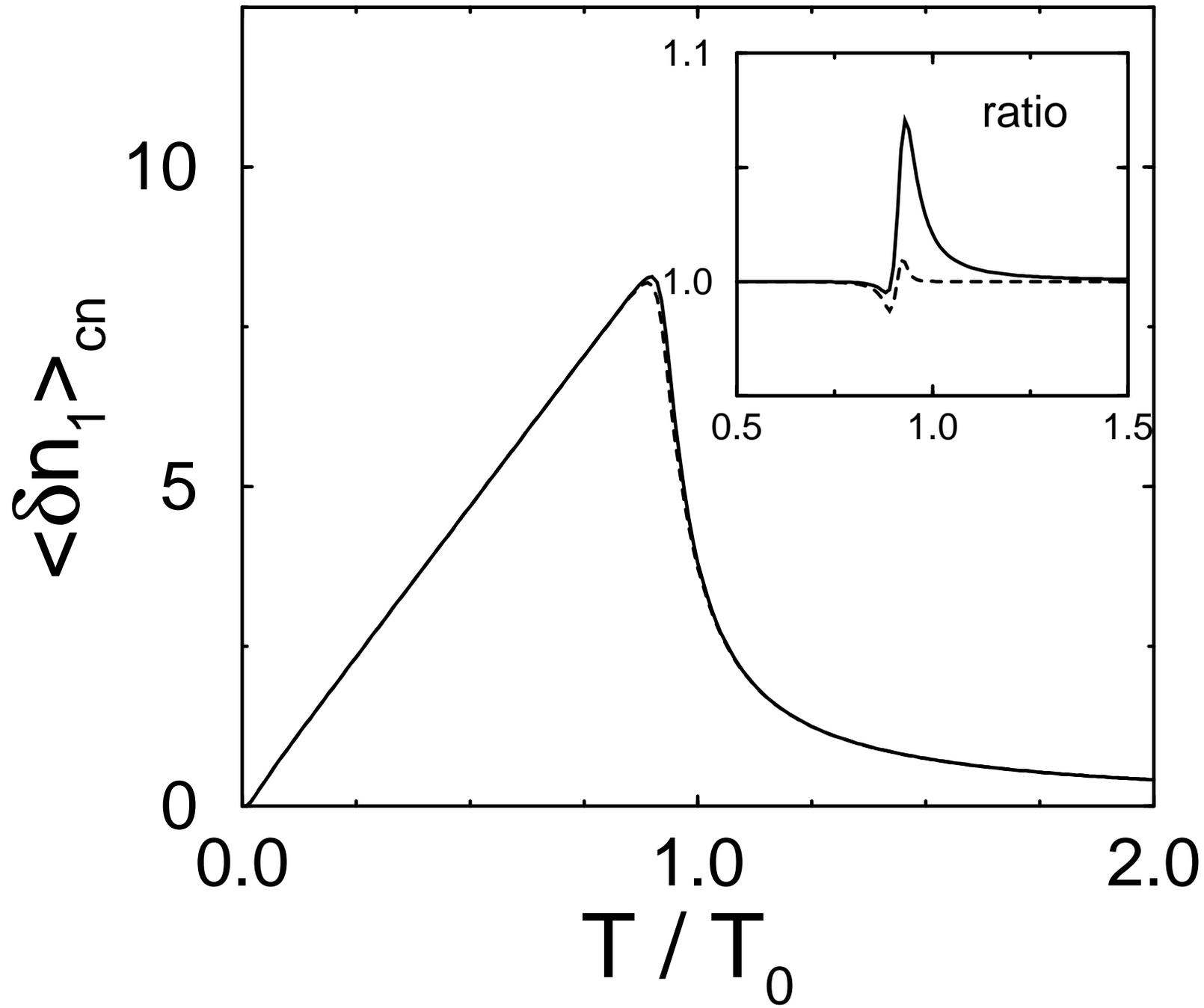
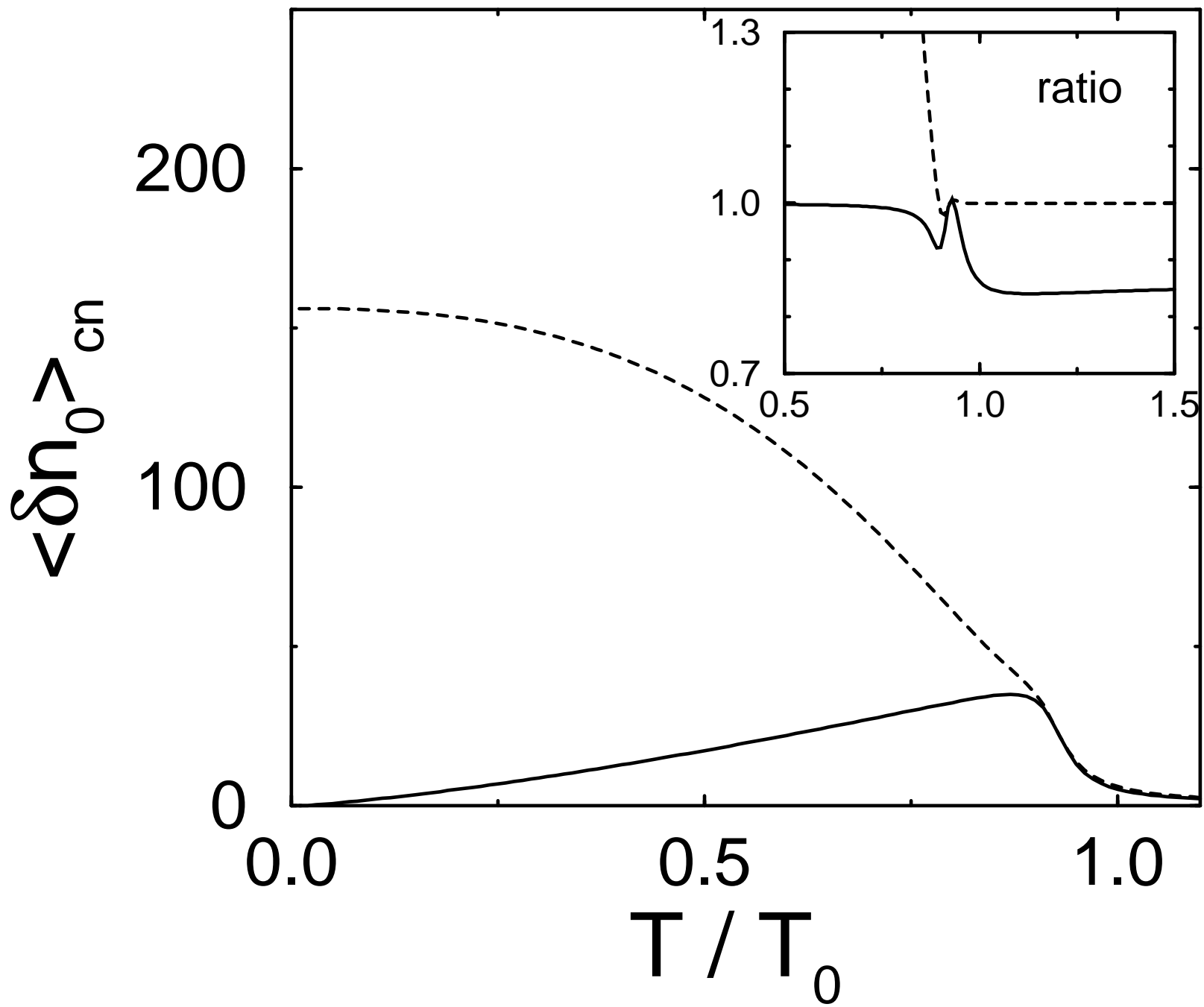
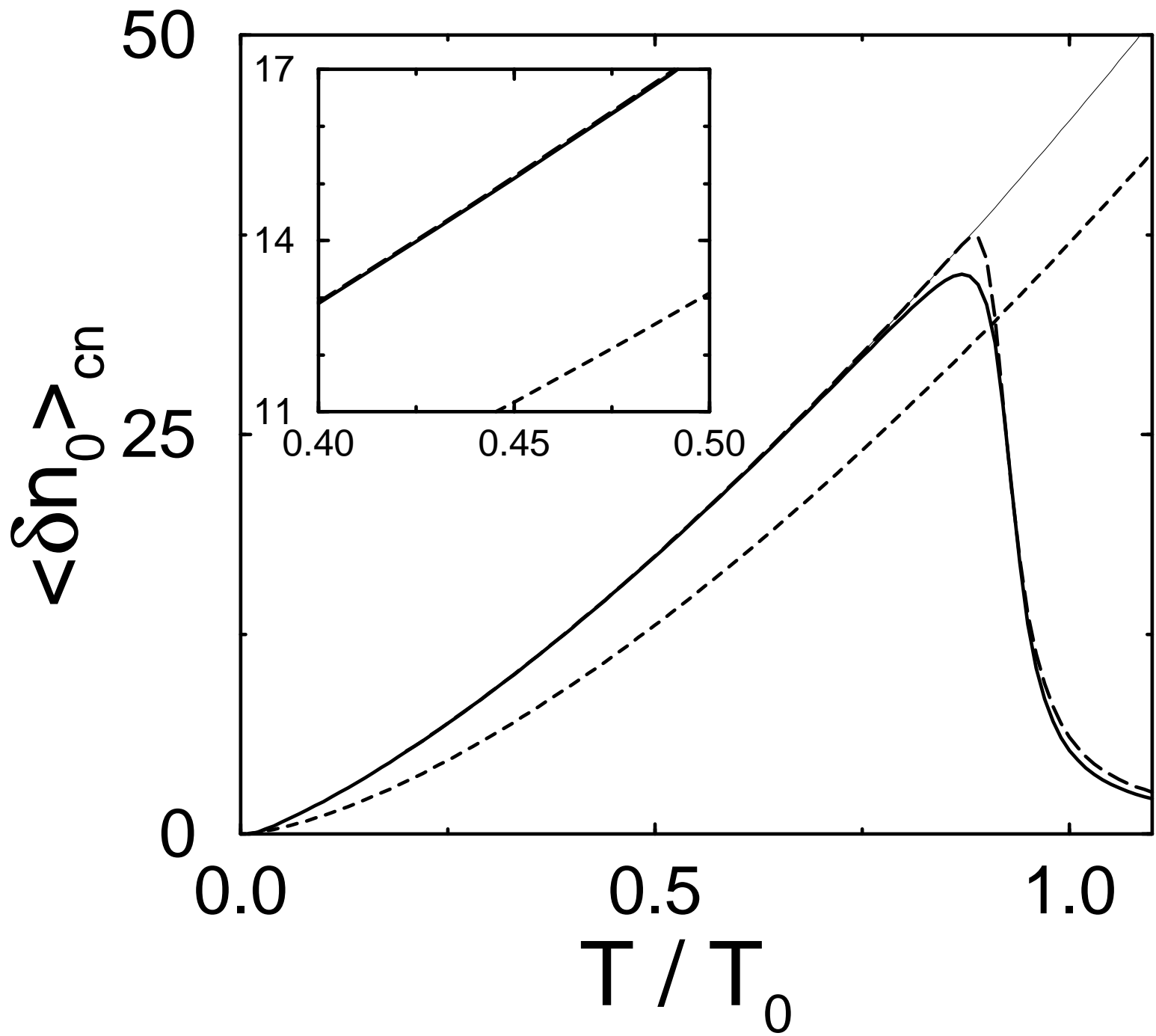
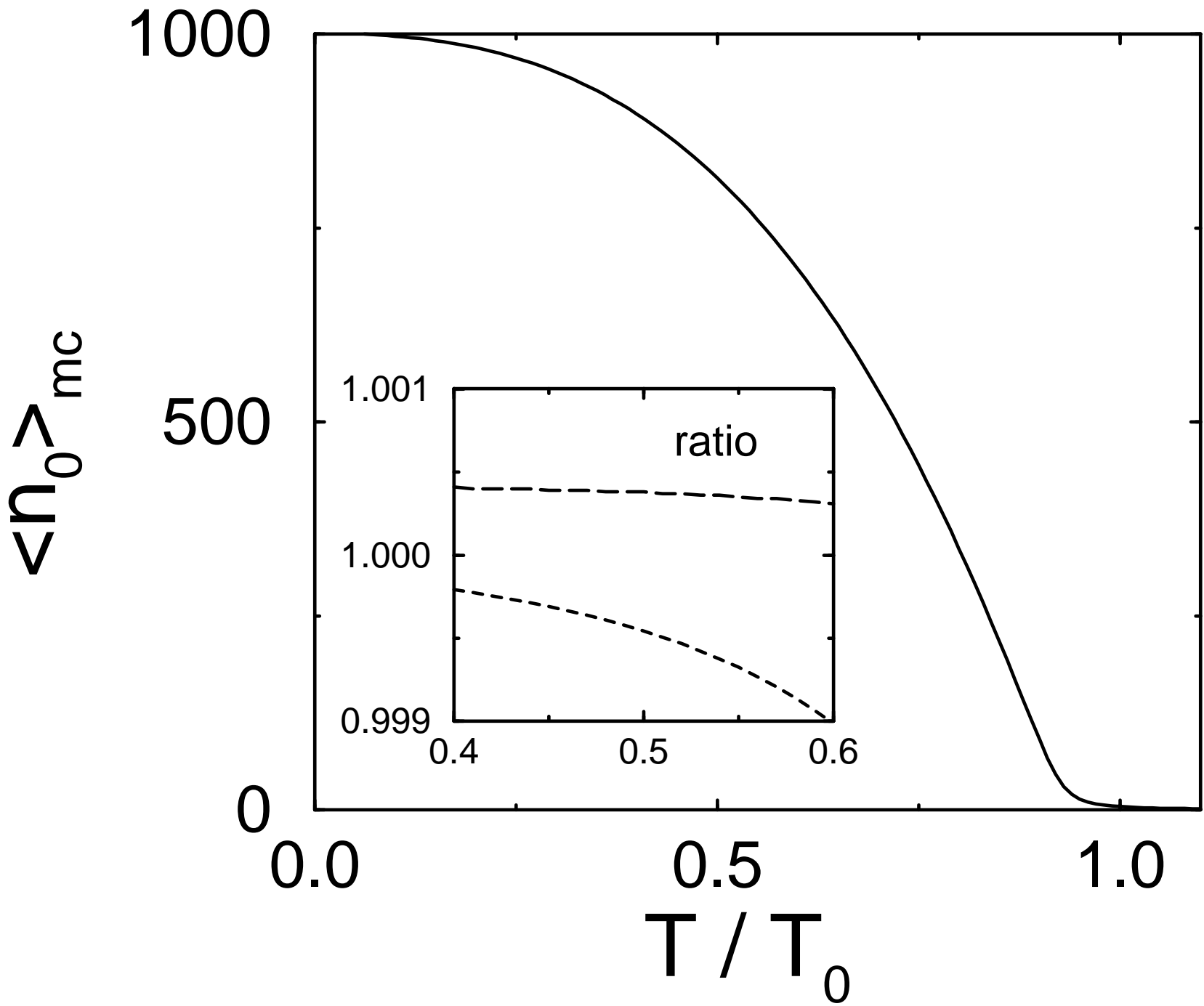
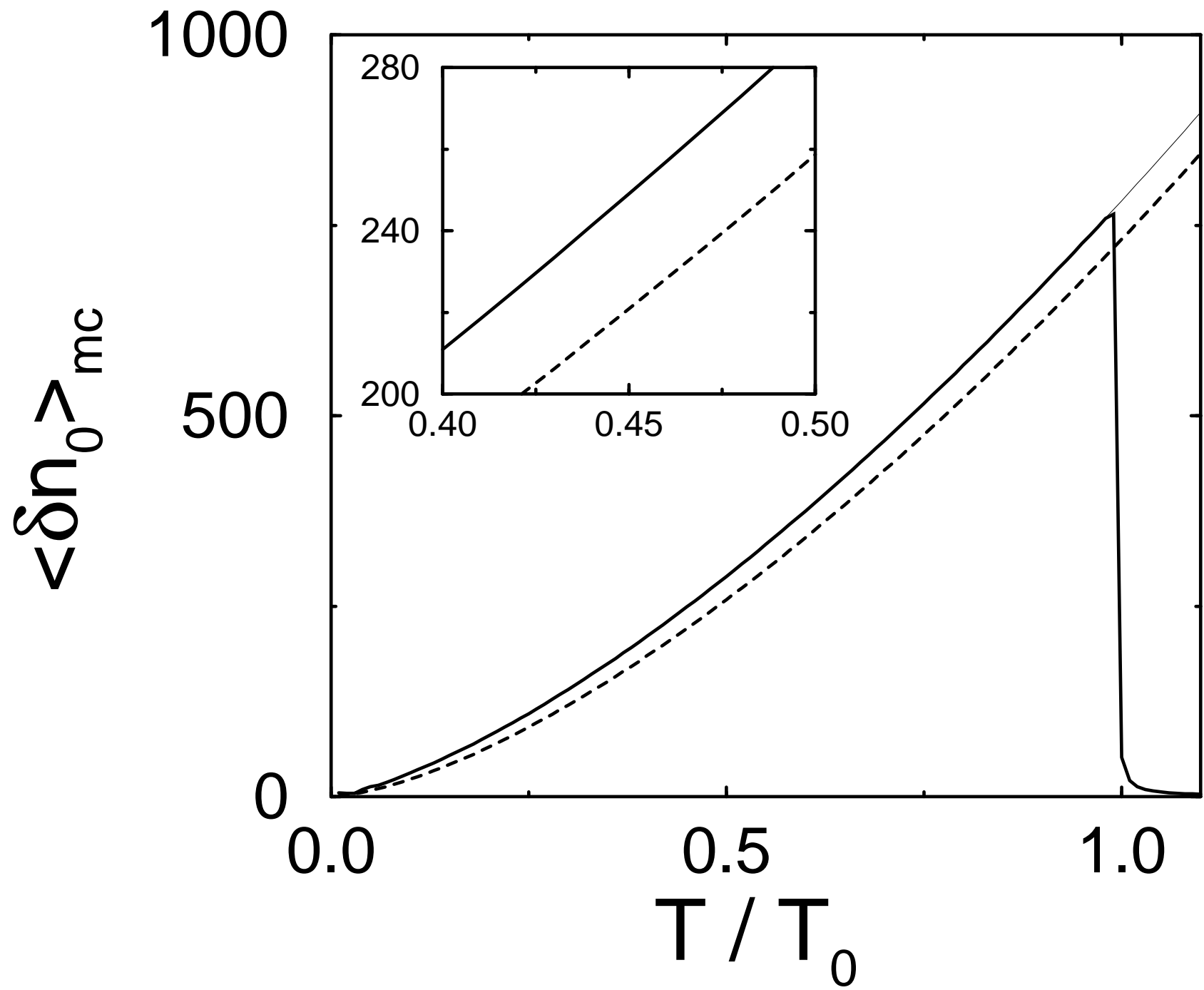
\end{document}